\documentclass[letterpaper]{article} 
\usepackage{aaai2026}  
\usepackage{times}  
\usepackage{helvet}  
\usepackage{courier}  
\usepackage[hyphens]{url}  
\usepackage{graphicx} 
\usepackage{natbib}  
\usepackage{caption} 
\usepackage{algorithm}
\usepackage{algorithmic}
\usepackage{comment}
\usepackage{subcaption}
\usepackage{graphicx}
\usepackage{threeparttable}
\usepackage{booktabs}
\usepackage{tabularx}

\usepackage{newfloat}
\usepackage{float}
\usepackage{listings}
\usepackage{tablefootnote}

\DeclareCaptionStyle{ruled}{labelfont=normalfont,labelsep=colon,strut=off} 
\floatstyle{ruled}
\newfloat{listing}{tb}{lst}{}
\floatname{listing}{Listing}
\usepackage{xcolor}
\usepackage{longtable}
\usepackage[breakable,listings]{tcolorbox}
\usepackage{array}

\newtcolorbox{longlisting}[2][]{
  breakable,                 
  title={#2},
  listing only,
  #1
}

\title{NOVAID: \textbf{N}atural-language \textbf{O}bservability \textbf{V}isualization \textbf{A}ssistant for \textbf{I}TOps \textbf{D}ashboard Widget Generation}
\author {
    Pratik Mishra\textsuperscript{\rm 1},
    Caner Gözübüyük\textsuperscript{\rm 2},
    Seema Nagar\textsuperscript{\rm 1},
    Prateeti Mohapatra\textsuperscript{\rm 1},
    Raya Wittich\textsuperscript{\rm 2},
    Arthur De Magalhaes\textsuperscript{\rm 2}
}
\affiliations {
    \textsuperscript{\rm 1}IBM Research\\
    \textsuperscript{\rm 2}IBM Software\\
    \{pratik-mishra, caner.goez\}@ibm.com, \{senagar3, pramoh01\}@in.ibm.com, raya.wittich@de.ibm.com, arthurdm@ca.ibm.com
}

\begin{document}

\maketitle

\begin{abstract}

Manual creation of IT monitoring dashboard widgets is slow, error-prone, and a barrier for both novice and expert users. We present NOVAID, an interactive chatbot that leverages Large Language Models (LLMs) to generate IT monitoring widgets directly from natural language queries. Unlike general natural language–to-visualization tools, NOVAID addresses IT operations–specific challenges: specialized widget types like SLO charts, dynamic API-driven data retrieval, and complex contextual filters. The system combines a domain-aware semantic parser, fuzzy entity matching, and schema completion to produce standardized widget JSON specifications. An interactive clarification loop ensures accuracy in underspecified queries. On a curated dataset of $271$ realistic queries, NOVAID achieves promising accuracy (up to $94.10\%$ in metric extraction) across multiple LLMs.
A user study with IT engineers yielded a System Usability Scale score of 74.2 for NOVAID, indicating good usability.
By bridging natural language intent with operational dashboards, NOVAID demonstrates clear potential and a path for deployment in enterprise ITOps monitoring platforms.

\end{abstract}



\section{Introduction}

The health and performance of modern IT systems are meticulously monitored by tracking various metrics and key performance indicators (KPIs) over time. This process, known as IT Operations (ITOps), involves a complex domain of data, which is often time-varying in nature and exposed through dynamic APIs. ITOps monitoring tools, such as Splunk~\cite{splunk_dashboard}, Datadog~\cite{datadog_dashboard}, and Instana~\cite{instana}, provide a means for site reliability engineers (SREs), developers, and other business professionals to track and assess system health through dashboards. While predefined IT dashboards are useful, they do not always cater to the specific needs of individual users. To address this, many existing tools ~\cite{instana_custom_dashboards, datadog_custom_dashboards, grafana_dashboard} offer the flexibility of creating custom dashboards that can be tailored to users' specific requirements by allowing them to add their desired widgets. These dashboards are crucial for better observability, offering users a centralized view of system health and performance while also enabling personalized visualization.

However, the manual process of creating and configuring individual widgets for these custom dashboards is often a tedious and time-consuming task for both new and experienced users, and it can be prone to errors. The complexity of the widget creation flow also extends the onboarding time for new users~\cite{elshehaly2020qualdash, rossi2024modeldriven}. This challenge is compounded by ITOps domain-specific issues, such as specialized widget types (e.g., Service Level Objective charts), complex data access patterns through dynamic APIs, and the need to handle nuanced, context-rich queries with conditional filters.

We propose NOVAID (\textbf{N}atural-language \textbf{O}bservability \textbf{V}isualization \textbf{A}ssistant for \textbf{I}TOps
\textbf{D}ashboards), an interactive chatbot powered by large language models (LLMs), to streamline the dashboard widget creation process. Our tool addresses the limitations of existing solutions by introducing a domain-specific, schema-aware approach to generate IT monitoring dashboard widgets from natural language queries. The tool parses user intent, extracts key elements, and interactively completes vague or incomplete queries through a multi-turn conversation, allowing users to quickly assemble custom dashboards without needing to understand the complex manual creation flow. The key contributions of our work are: (1) A novel AI application that tackles the significant and underexplored problem of natural language-to-dashboard widget generation in the ITOps domain. (2) A hybrid LLM-based architecture that combines an LLM-powered semantic parser with fuzzy entity matching to ensure precision and adherence to specialized IT monitoring data schemas. (3) An evaluation on a curated dataset of realistic ITOps queries and a user study with IT professionals, providing evidence of the system's effectiveness and its potential for real-world deployment.

\section{Related Work}

Natural Language to Visualization (NL2VIS)~\cite{luo2021nvbench,srinivasan2021collecting,wang2022towards,zhang2024natural} has attracted extensive attention, spawning both research prototypes and commercial products. Early systems~\cite{gao2015datatone, setlur2016eviza} pioneered mixed-initiative disambiguation and rapid statistical summarization. Toolkits like NL4DV~\cite{narechania2020nl4dv} provide end-to-end pipelines for parsing NL, inferring chart specifications, and rendering Vega-Lite graphics. Commercial BI platforms have incorporated NL interfaces at scale~\cite{jones2014communicating, knight2022microsoft}, allowing non-technical users to query enterprise datasets in plain English. Recent approaches~\cite{chen2024viseval,maddigan2023chat2vis} leverage LLMs to interpret user queries and generate visualization code conditioned on a database schema. There is emerging work on end-to-end dashboard generation~\cite{shi2020calliope, deng2022dashbot, srinivasan2023bolt, dibia2023lida} that can synthesize entire multi-chart dashboards from a single NL utterance.

However, existing NL2VIS solutions fall short when applied to the domain of ITOps and monitoring dashboards. These approaches—whether based on text-to-SQL translation or code generation for data retrieval and visualization—are not directly applicable to IT monitoring dashboards. IT monitoring platforms expose data through dynamic APIs. In this domain, visualization creation typically involves constructing a widget schema that defines the visual encoding, specifies parameters for API-based data retrieval, and determines the required time duration, reflecting the inherently time-series nature of IT data.

The following domain-specific challenges render existing NL2VIS techniques unsuitable for IT monitoring contexts:

\textbf{1. Domain-Specific Widget Types:} ITOps dashboards include specialized visuals like Service Level Objective (SLO) charts, which capture reliability metrics not commonly supported by general-purpose visualization libraries.

\textbf{2. Complex Data Access Patterns:} Data is accessed through APIs with dynamic parameter requirements. Even with fixed endpoints, constructing the correct query parameters—such as service names, application IDs, or metric types—demands precise interpretation of user intent.

\textbf{3. Conditional and Contextual Filters:} IT professionals pose nuanced, context-rich queries. For example:
    \begin{quote}
        Show me the average latency over time for all HTTP calls made to the \texttt{catalogue} service, as well as the total number of calls for each call type in the \texttt{otel-shop} application.
    \end{quote}
Such queries require identifying entities (e.g., \texttt{catalogue} service, \texttt{otel-shop} application), relevant filters (e.g., HTTP calls), and aggregations across different scopes (e.g., average latency vs. total calls by call type).

\section{NOVAID System Overview}

\begin{figure*}[!htp]
    \centering
    \includegraphics[width=\textwidth]{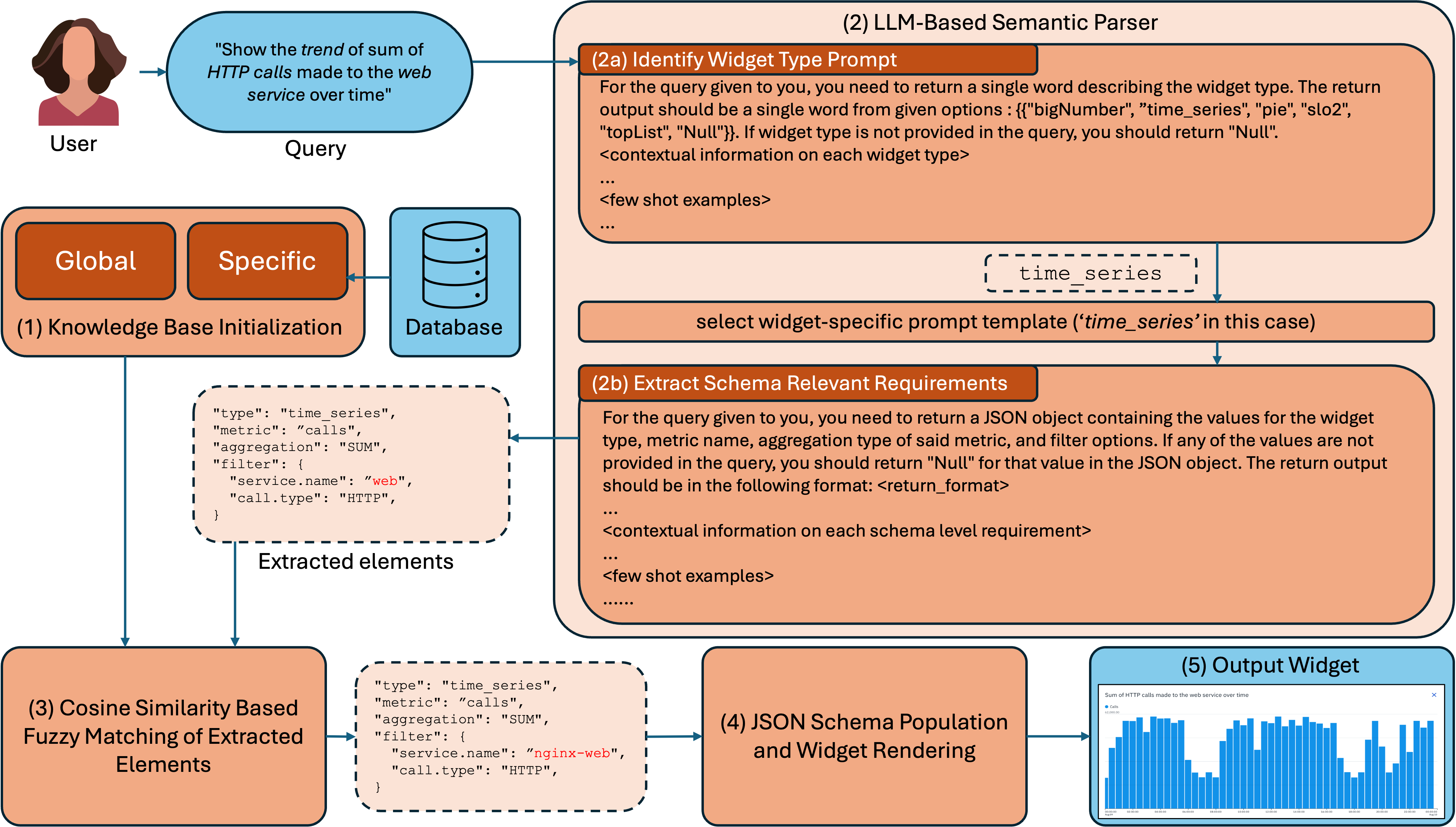} 
    \caption{Framework of our proposed method of NOVAID for generating widgets for custom dashboards using LLMs}
    \label{fig:system-overview}
\end{figure*}

NOVAID is designed to bridge the gap between natural language queries and fully functional, context-aware IT monitoring data visualizations. The system operates through a modular pipeline comprising tightly integrated components, each contributing to a seamless user experience—from understanding intent to rendering interactive dashboards. The process begins at a high level with the offline construction of a \textbf{knowledge base}, capturing global and instance-specific metadata. This foundation enables the subsequent inference of \textbf{key elements} from user queries.

To account for variations in terminology, wrong input, or incomplete input, we apply a robust \textbf{fuzzy matching} mechanism to align extracted elements with known entities in the system. Once these components are verified and completed, the system proceeds to \textbf{populate a structured JSON schema} which serves as the blueprint for rendering the desired widget. 


Figure~\ref{fig:system-overview} shows the components and the interactions. Each of these stages is elaborated in the subsections that follow, demonstrating how the system transitions from user intent to a rendered, data-rich visual with minimal friction and maximum semantic alignment.

\subsection{Knowledge Base Initialization}

Our system relies on an online initialization process to construct a structured knowledge base that encapsulates essential domain and system-specific metadata. This step is critical for enabling consistent and intelligent behavior during system operation. The knowledge base comprises two distinct categories of information:

\begin{itemize}
    \item \textbf{Type I: Global Knowledge.} This includes information that is agnostic to specific system instances and remains consistent across deployments. Examples include the set of supported widget types, metric names along with their allowed aggregation functions, and commonly used instance-invariant filter parameters. Since this data is static and universal, it is precomputed and stored as part of the system’s embedded configuration.
    
    \item \textbf{Type II: Instance-Specific Knowledge.} This captures dynamic, deployment-dependent information that varies across different instances of the system. Examples include available service names, application identifiers, and endpoint names — entities that are tightly coupled with the runtime environment. As this data cannot be predefined, the system retrieves it at runtime via RESTful API queries to the target environment upon deployment.
\end{itemize}


\subsection{Semantic Parser}
\label{key-elements-extraction}

Given a user query, our system extracts or infers (if the widget type is not mentioned in the query) the key elements required to construct a visualization widget. The most essential piece of information is the \textbf{widget type}, which determines how the data should be presented. A single widget can incorporate information from multiple data sources. For instance, a time series plot can display various series, each corresponding to a distinct data source.

A \textbf{data source} is defined as the combination of a \textit{metric name}, an \textit{aggregation function}, a set of \textit{filter options}, and a \textit{grouping criterion}. Filter options may range from simple attribute constraints to complex logical expressions.  

For instance, consider the query: \textit{``Show the mean latency of the catalogue service in the robot-shop application for HTTP calls, restricted to the top (bottom) five endpoints"} 

This request translates into the following composite filter:  

\begin{quote}
\texttt{application.name = "robot-shop" AND service.name = "catalogue" AND call.type = "HTTP"}
\end{quote}

The grouping criterion specifies how the system should partition results and the number of outputs to return. In this example, the corresponding grouping criterion is:  

\begin{quote}
\texttt{groupByTag = endpoint.name, maxResults = 5, direction = DESC}
\end{quote}

We use a \textbf{two-pass process} to infer both the widget type and its corresponding data sources, motivated by the need to disambiguate user intent early and handle widget-specific schema differences. 

\begin{itemize}
    \item \textbf{First Pass – Widget Type Inference:}  
    The LLM identifies the user's intent by selecting the appropriate widget type from a predefined knowledge base. This lightweight step provides essential context for the next phase.

    \item \textbf{Second Pass – Data Source Extraction:}  
    With the widget type known, the LLM extracts the specific data sources required. This separation is necessary because data schema requirements vary—e.g., an \textit{SLO widget} may need just a name, while a \textit{time series widget} may require multiple metrics.

\end{itemize}












\subsection{Fuzzy Matching of Extracted Elements}

The extracted key elements from the user query may not always exactly match entries in the system's knowledge base. Users may have incomplete names, abbreviations, or slightly incorrect names. To handle this, we employ a \textbf{fuzzy matching strategy} based on \textit{cosine similarity} between the extracted term and known values in the knowledge base. An \textbf{exact match} occurs if the cosine similarity score exceeds a defined threshold $th$, which is empirically decided. Based on the number and confidence of matches, we apply the following strategies:

\subsubsection*{1. Single High-Confidence Match (Auto-Correction)}
If there is only one candidate in the knowledge base whose similarity exceeds the threshold, we automatically replace the extracted value with this match.

\begin{itemize}
    \item \textbf{Example:} The user refers to \texttt{qotd-service}, but the knowledge base contains only \texttt{qotd-web service}. As the similarity is high and unique, the system replaces \texttt{qotd-service} with \texttt{qotd-web service}.
\end{itemize}

\subsubsection*{2. Multiple Similar Matches (User Disambiguation)}
If multiple candidate values exceed the similarity threshold, we present them to the user as a list of suggestions. The user is then prompted to make a selection.

\begin{itemize}
    \item \textbf{Example:} The user mentions \texttt{robot-shop service}, which does not exist in the knowledge base. However, two similar services, namely \texttt{robot-shop catalogue service} and \texttt{robot-shop shipping service}, are found. These are displayed to the user for manual selection.
\end{itemize}

\subsubsection*{3. Missing Key Element (Prompt for Completion)}\label{sub:missing} In cases where the user query omits an essential element, such as an aggregation method or specific filter value, we proactively detect the omission and prompt the user to choose from allowed values based on context.

\begin{itemize}
    \item \textbf{Example:} In the query, \textit{``I want to see the latency of the robot-shop application"}, no aggregation method is specified. The system detects this and prompts the user to select an appropriate aggregation (e.g., \texttt{mean}, \texttt{p95}, etc.) based on available options for the \texttt{latency} metric.
\end{itemize}



This fuzzy matching process increases system robustness to natural language variations and reduces the need for exact terminology, thereby improving usability for non-expert users.

\subsection{JSON Schema Population and Widget Rendering}

To enable dynamic dashboard generation, our system automatically constructs a JSON schema specific to each widget type by incorporating key parameters extracted from the preceding inference process, along with the user's selected time duration. This structured schema serves a dual purpose: it instructs the front-end on how to construct REST API requests for data retrieval, and it specifies the configuration required for rendering the corresponding widget. The JSON schema defines metadata, including the widget type and a configuration object. The configuration object specifies rendering options (e.g., line chart vs. bar chart), formatting preferences, and data source specifications. For SLO widgets, the data source is represented by the SLO name, whereas other widget types may reference multiple data sources.

\section{Experiments and Results}



\renewcommand{\thefootnote}{\arabic{footnote}} 


\begin{table*}[!htp]
\centering
\begin{threeparttable}
\renewcommand{\arraystretch}{1.3} 
\begin{tabular}{|>{\centering\arraybackslash}m{5cm}|c|c|c|c|c|c|}
\hline
\textbf{Large Language Model} & \multicolumn{6}{c|}{\textbf{Accuracy (\%)}} \\
\cline{2-7}
& \textbf{Widget Type} & \textbf{Metric} & \textbf{Aggregation} & \textbf{Tag Filter} & \textbf{Grouping} & \textbf{Overall} \\
\hline
\textbf{Granite-3-3-8b}\tnote{\dag} & 86.72 & 92.25 & 92.62 & 75.65 & \textbf{64.58} & 54.24 \\
\hline
\textbf{Llama-3-3-70b}\tnote{\dag} & \textbf{91.51} & 92.62 & 92.99 & 76.01 & \textbf{64.58} & 58.67 \\
\hline
\textbf{Llama-4-Maverick-17b-128e}\tnote{\dag} & 89.30 & \textbf{94.10} & \textbf{97.05} & \textbf{80.07} & 56.25 & 64.21 \\
\hline
\textbf{Mistral-Small-3.1-24b}\tnote{\dag} & 89.30 & \textbf{93.73} & \textbf{96.68} & \textbf{79.70} & 56.25 & 63.47 \\
\hline
\textbf{gpt-oss-20b} & 88.93 & 92.99 & 95.20 & 76.01 & 60.42 & 60.52 \\
\hline
\textbf{gpt-oss-120b} & \textbf{91.51} & 92.62 & 92.99 & 76.01 & \textbf{64.58} & \textbf{67.93} \\
\hline
\end{tabular}
\caption{Performance of various Large Language Models on Widget Generation}
\label{tab:llm_performance}
\begin{tablenotes}
\item[\dag] \textit{Instruction-tuned model was used.}
\end{tablenotes}
\end{threeparttable}
\end{table*}

\begin{table}[!htp]
\centering
\renewcommand{\arraystretch}{1.2}
\begin{tabular}{|l|c|c|c|}
\hline
\textbf{Model} & \textbf{Zero-Shot} & \textbf{Few-Shot} & $\mathbf{\Delta}$ \\
\hline
\textbf{Llama-3-3-70b\textsuperscript{\textdagger}} & 62.96\% & 67.41\% & \textbf{+4.45} \\
\textbf{Granite-3-3-8b\textsuperscript{\textdagger}} & 62.22\% & 73.88\% & \textbf{+11.66} \\
\hline
\end{tabular}
\caption{Tag Filter Expression accuracy improvement from Few-Shot prompting on a 242 query subset of the dataset}
\label{tab:fewshot_filter}
\end{table}

\subsection*{Dataset}
A major challenge in developing and evaluating NL interfaces for specialized domains is the lack of public datasets. To address this, we curated a custom dataset of $271$ realistic, single-turn queries. These queries were gathered from real-world scenarios and discussions with IT professionals, simulating the types of requests an end-user would make and factoring in situations where a user makes a mistake or misses out some information. This meant that the dataset included queries where one or more key elements are intentionally absent in order to assess the ``missing key element" functionality (discussed in System Overview~\ref{sub:missing} section). For such cases, the corresponding fields in the ground truth dataset were set to \textit{``null"}. 
The dataset is specifically structured to test the system's ability to extract and accurately map the following five key parameters to generate a valid widget JSON. \textbf{1. Widget Type:} The type of visualization requested (\textit{e.g., bigNumber, time\_series, topList}).
\textbf{2. Metric:} The specific performance metric to be displayed (\textit{e.g., calls, latency, erroneousCalls}).
\textbf{3. Aggregation:} The function to be applied to the metric \textit{(e.g., SUM, MEAN, PER\_SECOND)}.
\textbf{4. Tag Filter Expression:} Complex contextual filters based on key-value pairs (e.g., \texttt{{`service.name': `payment'}, \texttt{`call.type': `HTTP'}}).
\textbf{5. Grouping:} The entity by which the data should be grouped, where applicable (e.g., \texttt{`groupbyTag': `endpoint.name', `direction': `DESC', `maxResults': 10})
Of the total $271$ queries, $48$ explicitly required grouping, presenting a more complex challenge for the model.

\subsection{Evaluation Methodology}

We evaluated the system's performance by measuring the accuracy of each extracted field (\textit{widget type}, \textit{metric}, \textit{aggregation}, \textit{tag filter expression}, and \textit{grouping}) post knowledge-base cosine similarity comparison against the ground truth labels in our dataset. The primary metric for success is overall accuracy, defined as the percentage of queries where all five fields were extracted correctly, resulting in a fully valid and runnable widget JSON. This strict evaluation criterion ensures that a single error in any component invalidates the entire result, reflecting the practical requirements of generating production-ready dashboard configurations.


\subsection{Model Details and Configuration}
We tested our system with several open-source LLMs to demonstrate the robustness and independence of our pipeline from any single model. The LLMs evaluated were \textbf{Granite-3-3-8b-instruct}~\cite{granite_3.3_8b_instruct}, \textbf{Llama-3-3-70b-instruct}~\cite{llama3_70b}, \textbf{Llama-4-Maverick-17b-128e-instruct}~\cite{llama4_maverick_17b_128e_instruct}, \textbf{Mistral-Small-3.1-24B-Instruct}~\cite{mistral_small3_1_24b_instruct}, \textbf{gpt-oss-20b}, and \textbf{gpt-oss-120b}~\cite{gpt_oss}. Each model was given a format-tailored system prompt (Appendix Section A) and evaluated using a greedy decoding strategy with a maximum sequence length of $1024$ tokens, avoiding the randomness introduced by sampling methods like top-k or nucleus.


\subsection{Results}


Table~\ref{tab:llm_performance} summarizes the performance of various models on our evaluation bed. We initially tested zero-shot prompting to assess out-of-the-box model abilities, but the results were inconsistent. To improve robustness, we introduced few-shot prompting with \textbf{$15$ diverse examples} in each prompt chosen to maximize coverage across different parameters. As shown in Table~\ref{tab:fewshot_filter}, few-shot prompting improved accuracy and stability, leading us to adopt it for all subsequent evaluations.

All models tested achieved high accuracy in identifying core components of a query, with Metric and Aggregation accuracies consistently above $92\%$, reaching a peak of $97.05\%$ with the Llama-4-Maverick-17b-128e-instruct. The high accuracy in these fields confirms the viability of our domain-aware semantic parsing and LLM-based architecture for this complex task. The overall accuracy results, which are a highly conservative measure of performance, show that our system can successfully generate a complete and valid widget JSON for a significant portion of queries. Gpt-oss-120b achieved the highest overall accuracy at $67.0\%$ (however, not a significant improvement when compared with granite-3-3-8b at p-value $< 0.05$), indicating its best ability to handle all five parameters simultaneously. A closer look at the data reveals that the primary challenge lies in extracting complex Tag Filter Expressions and, in particular, Grouping parameters. While Tag Filter Expression accuracy reached a respectable $80.07\%$, the Grouping accuracy varied, peaking at $64.58\%$. This is consistent with the nature of the task; grouping is often underspecified or implicit in a query, requiring more advanced reasoning to infer the user's intent. For example, a user asking for \textit{``the services with the highest latency"} implies grouping by \texttt{service.name}, but the grouping field itself is not explicitly stated (see Appendix B for detailed error analysis). This is where our proposed multi-turn clarification loop becomes essential for practical deployment, as it addresses the ambiguities that a single-turn parsing approach cannot resolve.

\section{Deployment}

\begin{figure}[!htp]
\centering
\includegraphics[width=\linewidth]{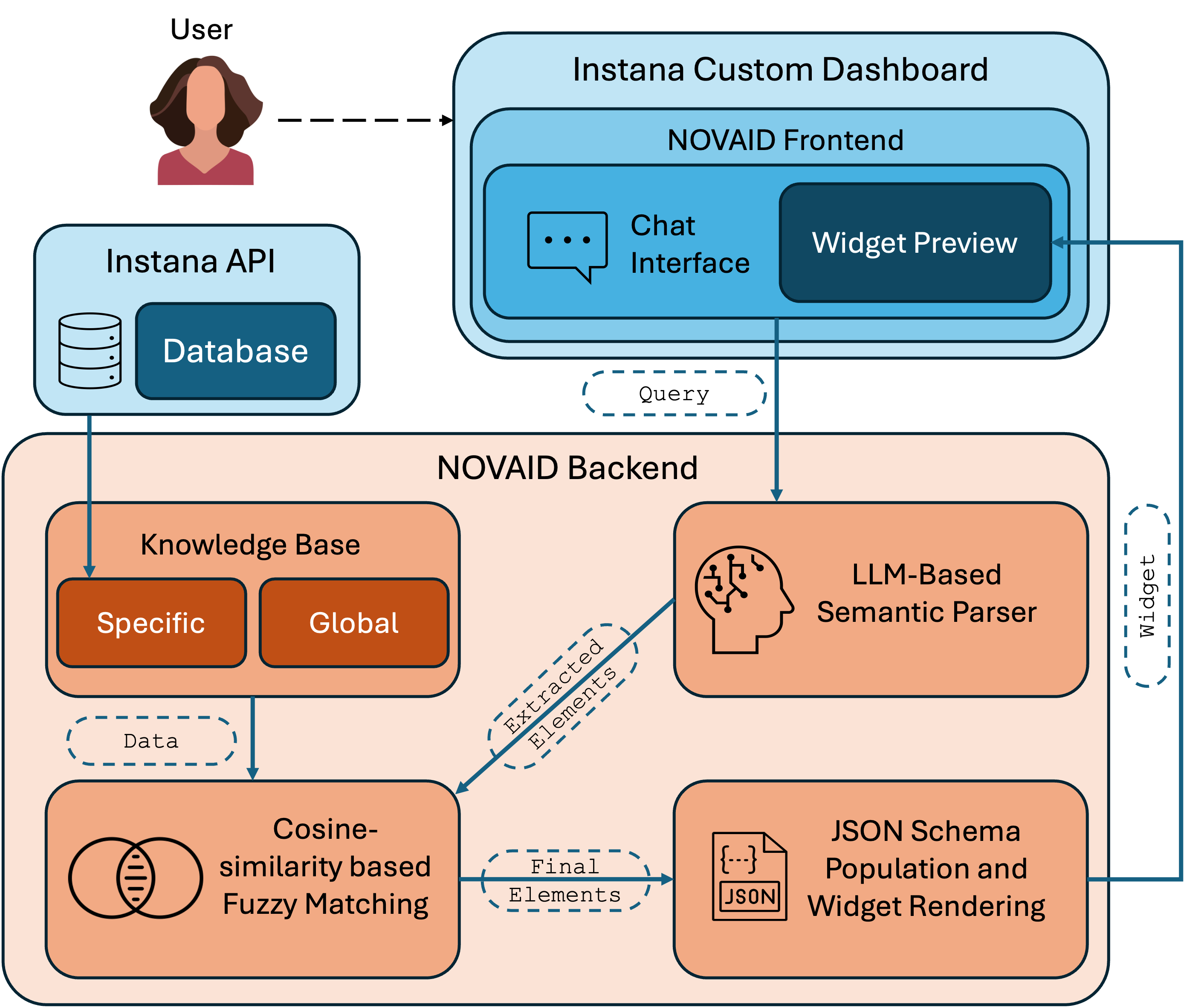}
\caption{NOVAID Deployment and Integration with Instana}
\label{fig:system-architecture}
\end{figure}

\begin{figure}[!htp]
\centering
\includegraphics[width=\linewidth]{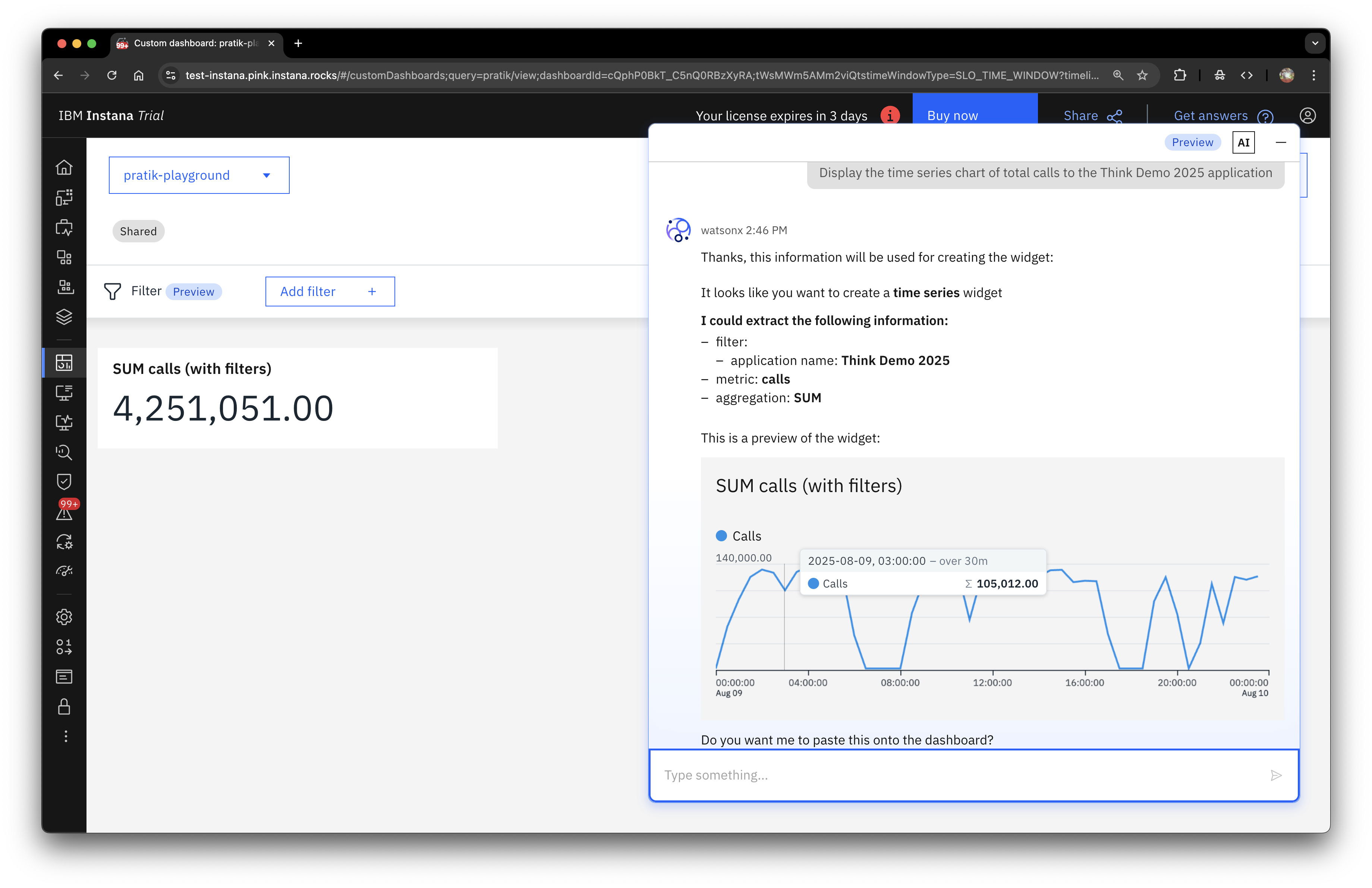}
\caption{NOVAID Frontend with query and widget preview}
\label{fig:frontend}
\end{figure}

NOVAID is embedded as an emerging application within the \textbf{Instana platform}~\cite{instana}, augmenting the traditional custom dashboard workflow with natural language interaction. Rather than requiring users to manually configure widgets through multi-step forms that are traditionally present and used in Instana Custom Dashboard's widget creation tool, NOVAID enables conversational widget authoring, lowering the barrier for SREs and developers to track and assess system health from monitoring data. Figure~\ref{fig:system-architecture} goes over NOVAID’s architecture and integration into the Instana ecosystem. The system consists of two main components: a \textbf{front-end interface} with chat and interactive widget preview (see Figure~\ref{fig:frontend}), and a \textbf{backend service} for query translation. 

\subsection{NOVAID Frontend}
The front end is deployed on Instana’s multi-region Kubernetes infrastructure, ensuring high availability and smooth integration with existing workflows.
 
\textbf{The chat interface} is the primary entry point for users, embedded directly within the Instana dashboard. Users express their requests in natural language—such as filters, aggregation strategies, or visualization preferences—and receive our tool's interpretations within the chat window. The interface supports multi-turn conversations, prompting for missing parameters and offering clarification options when user input is ambiguous or is missing key elements.

\textbf{The widget preview} renders an immediate interactive visualization inside the chat pane once NOVAID resolves a user’s intent. Users can refine the preview through follow-up interactions, and once satisfied, confirm its addition to the main dashboard pane. This reduces trial-and-error in the manual widget builder and accelerates dashboard authoring


\subsection{NOVAID Backend}


The backend implements the natural-language-to-visualization pipeline, connecting the chat interface with Instana’s monitoring services. Its responsibilities include (\textit{i) intent parsing and parameter resolution}, \textit{(ii) retrieving metadata and metrics from Instana APIs}, and \textit{(iii) generating specifications for widget previews and confirmed dashboard widget}s. RESTful APIs expose this functionality for dynamic and automated integration. The backend runs in a per-tenant, containerized deployment on Instana’s cloud infrastructure, enforcing isolation and authentication. Its modular microservices design allows incremental rollout of additional widget types (e.g. Histogram, Table) without disruption.

\subsection{Operational Status and Rollout}
NOVAID has been deployed (challenges covered in Appendix C) in a test system for an internal user study for several months, demonstrating the potential of conversational workflows to reduce the dashboard customization efforts. Based on these evaluations, it is now being introduced as a public preview for Instana customers. IBM's LLM \textbf{Granite-3-3-8b-instruct} was chosen with its smaller size offering lower inference cost and latency compared to gpt-oss-120b, whose accuracy improvement is not statistically significant. 
Early users highlight the chat interface's value for intent clarification and the widget preview for rapid iteration, demonstrating both immediate utility and a clear trajectory to production readiness. Ongoing effort focuses on expanding supported widget types, improving disambiguation of domain-specific terminology, and validating scalability under production loads. Addressing these challenges forms the next step toward full-scale deployment.

\section{User Study}



To evaluate NOVAID's effectiveness and usability, we conducted a preliminary user study with IT support engineers from different domains (e.g., network, storage), including both experienced and novice users. Participants performed two tasks: manually creating widgets and using NOVAID for automatic creation, followed by a survey. They were encouraged to explore NOVAID beforehand and provided with task descriptions to guide their natural language queries. For each task, participants identified an IT fault and considered investigative information to aid remediation. Study data included widget correctness and survey responses. After completing the tasks, participants filled out a usability questionnaire and provided feedback. Selected user comments appear in Appendix E. Table~\ref{label:correctness} presents widget correctness, rated on a 5-star scale (5 stars = $100\%$). Correctness improved with each interaction, reflecting a learning curve.

\begin{table}[!htp]
\centering
\begin{tabular}{|c|c|c|}
    \hline
    \textbf{First Widget} & \textbf{Second Widget} & \textbf{Third Widget} \\
    \hline
    4 & 4 & 4 \\
    1 & 2 & 5 \\
    5 & 5 & 5 \\
    1 & 5 & 5 \\
    5 & 4 & 5 \\
    3 & 5 & 3 \\
    \hline
    \multicolumn{3}{|c|}{average correctness} \\
    \hline
    \textbf{3.17} & \textbf{4.17} & \textbf{4.5} \\
    \hline
\end{tabular}
\caption{Correctness of automatic widget creation}
\label{label:correctness}
\end{table}

The second survey section used standard usability benchmark questions from~\cite{SUS} to assess NOVAID. Table~\ref{label:questions} reports average scores from six participants. NOVAID achieved a System Usability Scale (SUS) score of $74.2$, exceeding the benchmark of $68$~\cite{Sauro2016} and indicating good usability. Responses were on a five-point Likert scale~\cite{albaum1997likert}, with odd items positive (higher = better) and even items negative (lower = better). Participants shared subjective feedback on their experience creating widgets. Overall sentiment was positive, with users finding NOVAID promising and easy to use.

\begin{table}[!htp]
\centering
\begin{tabular}{|p{7cm}|p{1cm}|}
    \hline
    \textbf{Statement} & \textbf{Average score} \\
    \hline
    1. I think that I would like to use this system frequently. & \textbf{3.67} \\
    \hline
    2. I found the system unnecessarily complex. &  \textbf{1.83} \\
    \hline
    3. I thought the system was easy to use. & \textbf{3.50} \\
    \hline
    4. I think that I would need the support of a technical person to be able to use this system. & \textbf{2.50} \\
    \hline
    5. I would imagine that most people would learn to use this system very quickly. & \textbf{4.50} \\
    \hline
    6. I thought there was too much inconsistency in this system. & \textbf{2.50} \\
    \hline
    7. I felt very confident using the system. & \textbf{3.83} \\
    \hline
    8. I found the system very cumbersome to use. & \textbf{2.00} \\
    
    \hline
    9. I found the various functions in this system were well integrated. & \textbf{4.33} \\
    \hline
    10. I needed to learn a lot of things before I could get going with this system. & \textbf{1.33} \\
    \hline
    
    \hline
\end{tabular}
\caption{Usability Study}
\label{label:questions}
\end{table}

\section{Conclusion and Future Work}

In this paper, we introduced a method for automatically generating widgets for custom dashboards in IT monitoring tools via a conversational interface. Our approach performed well on curated queries, and a study confirmed users found the system intuitive and useful. The implementation and evaluation of NOVAID demonstrate the feasibility of generating dashboard widgets for ITOps. For future work, we will address limitations identified by user study participants (Appendix E). First, we will improve key element extraction using dynamic few-shot learning. Second, we will refine prompts to reduce cost and model latency. Third, we aim to expand functionality by supporting more widget types and integrating additional data sources. Finally, we will enhance the user experience with error feedback and iterative widget refinement through natural language. Since NOVAID is ITOps tailored, it may not fully meet other domain-specific needs out-of-the-box, slightly limiting applicability.

\bibliography{template}

@book{Sauro2016,
  author    = {Jeff Sauro and James R. Lewis},
  title     = {Quantifying the User Experience: Practical Statistics for User Research},
  edition   = {2nd},
  year      = {2016},
  publisher = {Morgan Kaufmann},
  address   = {Boston, MA},
  isbn      = {978-0128023082}
}

@misc{instana,
  author = {Instana},
  title = {IBM Instana Observability},
  url = {\url{https://www.ibm.com/products/instana}},
  year = {2024},
}

@article{albaum1997likert,
  title={The Likert scale revisited},
  author={Albaum, Gerald},
  journal={Market Research Society. Journal.},
  volume={39},
  number={2},
  pages={1--21},
  year={1997},
  publisher={SAGE Publications Sage UK: London, England}
}

@misc{splunk_dashboard,
  author = {{Splunk Inc.}},
  title = {Getting Started with Dashboards in Splunk},
  year = {2024},
  url = {https://docs.splunk.com/Documentation/Splunk/latest/Viz/AboutDashboards}
}

@misc{datadog_dashboard,
  author = {{Datadog Inc.}},
  title = {Dashboards Overview - Datadog Documentation},
  year = {2024},
  url = {https://docs.datadoghq.com/dashboards/}
}

@misc{datadog_custom_dashboards,
  author = {{Datadog Inc.}},
  title = {Create and Manage Custom Dashboards},
  year = {2024},
  url = {https://docs.datadoghq.com/dashboards/configure/}
}

@misc{instana_custom_dashboards,
  author = {{Instana}},
  title = {Creating Custom Dashboards in Instana},
  year = {2024},
  url = {https://www.ibm.com/docs/en/instana-observability/current?topic=guides-custom-dashboards}
}

@misc{rossi2024modeldriven,
  author       = {Rossi, Maria Teresa and Tundo, Alessandro and Mariani, Leonardo},
  title        = {Towards Model-Driven Dashboard Generation for Systems-of-Systems},
  howpublished = {\url{https://arxiv.org/abs/2402.15257}},
  year         = {2024},
}

@article{elshehaly2020qualdash,
  title={QualDash: Adaptable generation of visualisation dashboards for healthcare quality improvement},
  author={Elshehaly, Mai and Randell, Rebecca and Brehmer, Matthew and McVey, Lynn and Alvarado, Natasha and Gale, Chris P and Ruddle, Roy A},
  journal={IEEE Transactions on Visualization and Computer Graphics},
  volume={27},
  number={2},
  pages={689--699},
  year={2020},
  publisher={IEEE}
}

@article{deng2022dashbot,
  title={DashBot: Insight-driven dashboard generation based on deep reinforcement learning},
  author={Deng, Dazhen and Wu, Aoyu and Qu, Huamin and Wu, Yingcai},
  journal={IEEE Transactions on Visualization and Computer Graphics},
  volume={29},
  number={1},
  pages={690--700},
  year={2022},
  publisher={IEEE}
}

@article{shi2020calliope,
  title={Calliope: Automatic visual data story generation from a spreadsheet},
  author={Shi, Danqing and Xu, Xinyue and Sun, Fuling and Shi, Yang and Cao, Nan},
  journal={IEEE Transactions on Visualization and Computer Graphics},
  volume={27},
  number={2},
  pages={453--463},
  year={2020},
  publisher={IEEE}
}

@article{narechania2020nl4dv,
  title={NL4DV: A toolkit for generating analytic specifications for data visualization from natural language queries},
  author={Narechania, Arpit and Srinivasan, Arjun and Stasko, John},
  journal={IEEE Transactions on Visualization and Computer Graphics},
  volume={27},
  number={2},
  pages={369--379},
  year={2020},
  publisher={IEEE}
}

@book{jones2014communicating,
  title={Communicating data with Tableau: designing, developing, and delivering data visualizations},
  author={Jones, Ben},
  year={2014},
  publisher={" O'Reilly Media, Inc."}
}

@book{knight2022microsoft,
  title={Microsoft Power BI Quick Start Guide: The ultimate beginner's guide to data modeling, visualization, digital storytelling, and more},
  author={Knight, Devin and Ostrowsky, Erin and Pearson, Mitchell and Schacht, Bradley},
  year={2022},
  publisher={Packt Publishing Ltd}
}

@inproceedings{gao2015datatone,
  author       = {Gao, Tong and Dontcheva, Mira and Adar, Eytan and Karahalios, Karrie},
  title        = {DataTone: Managing Ambiguity in Natural Language Interfaces for Data Visualization},
  booktitle    = {Proceedings of the 28th Annual ACM Symposium on User Interface Software and Technology (UIST ’15)},
  year         = {2015},
  pages        = {117--126},
  publisher    = {ACM},
  address      = {New York, NY, USA},
  doi          = {10.1145/2807442.2807448},
}

@inproceedings{srinivasan2023bolt,
  author       = {Srinivasan, Arjun and Stasko, John},
  title        = {BOLT: A Natural Language Interface for Dashboard Authoring},
  booktitle    = {EuroVis 2023 – Short Papers},
  editor       = {Hoellt, Thomas and Aigner, Wolfgang and Wang, Bei},
  publisher    = {Eurographics},
  year         = {2023},
  pages        = {7--11},
  doi          = {10.2312/evs.20231035},
}

@inproceedings{dibia2023lida,
  author       = {Dibia, Victor},
  title        = {{LIDA}: A Tool for Automatic Generation of Grammar-Agnostic Visualizations and Infographics Using Large Language Models},
  booktitle    = {Proceedings of the 61st Annual Meeting of the Association for Computational Linguistics (Volume 3: System Demonstrations)},
  editor       = {Bollegala, Danushka and Huang, Ruihong and Ritter, Alan},
  pages        = {113--126},
  month        = jul,
  year         = {2023},
  publisher    = {Association for Computational Linguistics},
  doi          = {10.18653/v1/2023.acl-demo.11},
  url          = {https://aclanthology.org/2023.acl-demo.11/},
}

@inproceedings{setlur2016eviza,
  author       = {Setlur, Vidya and Battersby, Sarah E. and Tory, Melanie and Gossweiler, Rich and Chang, Angel X.},
  title        = {Eviza: A Natural Language Interface for Visual Analysis},
  booktitle    = {Proceedings of the 29th Annual Symposium on User Interface Software and Technology (UIST ’16)},
  year         = {2016},
  pages        = {365--377},
  publisher    = {ACM},
  address      = {New York, NY, USA},
  doi          = {10.1145/2984511.2984582},
}

@article{zhang2024natural,
  title={Natural language interfaces for tabular data querying and visualization: A survey},
  author={Zhang, Weixu and Wang, Yifei and Song, Yuanfeng and Wei, Victor Junqiu and Tian, Yuxing and Qi, Yiyan and Chan, Jonathan H and Wong, Raymond Chi-Wing and Yang, Haiqin},
  journal={IEEE Transactions on Knowledge and Data Engineering},
  year={2024},
  publisher={IEEE}
}

@article{chen2024viseval,
  title={Viseval: A benchmark for data visualization in the era of large language models},
  author={Chen, Nan and Zhang, Yuge and Xu, Jiahang and Ren, Kan and Yang, Yuqing},
  journal={IEEE Transactions on Visualization and Computer Graphics},
  year={2024},
  publisher={IEEE}
}

@article{maddigan2023chat2vis,
  title={Chat2vis: Generating data visualizations via natural language using chatgpt, codex and gpt-3 large language models},
  author={Maddigan, Paula and Susnjak, Teo},
  journal={Ieee Access},
  volume={11},
  pages={45181--45193},
  year={2023},
  publisher={Ieee}
}

@article{wang2022towards,
  title={Towards natural language-based visualization authoring},
  author={Wang, Yun and Hou, Zhitao and Shen, Leixian and Wu, Tongshuang and Wang, Jiaqi and Huang, He and Zhang, Haidong and Zhang, Dongmei},
  journal={IEEE Transactions on Visualization and Computer Graphics},
  volume={29},
  number={1},
  pages={1222--1232},
  year={2022},
  publisher={IEEE}
}

@article{luo2021nvbench,
  title={nvBench: A large-scale synthesized dataset for cross-domain natural language to visualization task},
  author={Luo, Yuyu and Tang, Jiawei and Li, Guoliang},
  journal={arXiv preprint arXiv:2112.12926},
  year={2021}
}

@inproceedings{srinivasan2021collecting,
  title={Collecting and characterizing natural language utterances for specifying data visualizations},
  author={Srinivasan, Arjun and Nyapathy, Nikhila and Lee, Bongshin and Drucker, Steven M and Stasko, John},
  booktitle={Proceedings of the 2021 CHI Conference on Human Factors in Computing Systems},
  pages={1--10},
  year={2021}
}

@article{sus,
author = {Lewis, James},
year = {2018},
month = {03},
pages = {1-14},
title = {The System Usability Scale: Past, Present, and Future},
journal = {International Journal of Human-Computer Interaction},
doi = {10.1080/10447318.2018.1455307}
}

@misc{grafana_dashboard,
  author = {{Grafana Labs}},
  title = {{Build your first dashboard}},
  howpublished = {\url{https://grafana.com/docs/grafana/latest/getting-started/build-first-dashboard/}},
  year = {2025}
}

@misc{granite_3.3_8b_instruct,
  author = {{Granite, IBM}},
  title = {{Granite-3.3-8B-Instruct} Model Card},
  howpublished = {\url{https://huggingface.co/ibm-granite/granite-3.3-8b-instruct}},
  year = {2025}
}

@misc{llama3_70b,
  author = {{Meta}},
  title = {{LLaMA 3 (70B)} Language Model},
  howpublished = {\url{https://www.llama.com/models/llama-3/}},
  year = {2024}
}

@misc{llama4_maverick_17b_128e_instruct,
  author       = {{Meta}},
  title        = {{Llama-4 Maverick 17B-128E-Instruct} Language Model},
  howpublished = {\url{https://huggingface.co/meta-llama/Llama-4-Maverick-17B-128E-Instruct}},
  year         = {2025}
}

@misc{mistral_small3_1_24b_instruct,
  author       = {{Mistral}},
  title        = {{Mistral-Small-3.1-24B-Instruct-2503} Language Model},
  howpublished = {\url{https://huggingface.co/mistralai/Mistral-Small-3.1-24B-Instruct-2503}},
  year         = {2025}
}

@misc{gpt_oss,
  author       = {{OpenAI}},
  title        = {{gpt-oss} Language Model},
  howpublished = {\url{https://openai.com/index/introducing-gpt-oss/}},
  year         = {2025}
}

\clearpage
\appendix
\renewcommand{\thesection}{\Alph{section}}
\renewcommand{\thefigure}{\Alph{section}\arabic{figure}}
\renewcommand{\thetable}{\Alph{section}\arabic{table}}
\renewcommand{\thesubsection}{\Alph{subsection}}
\setcounter{section}{0}
\setcounter{figure}{0}
\setcounter{table}{0}

\section*{Appendix}
\addcontentsline{toc}{section}{Appendix}

\subsection{A. Model Prompts}
\label{model_prompts}
This section outlines the prompts used for each model in widget type identification and extracting widget-specific information. We present the different prompts tested for every model. The model is composed of a system prompt followed by a prompt template tailored to that specific model. The specific prompts are the same for all models.

\begin{longlisting}{Prompt for Widget Type Identification}
\begin{lstlisting}
<|start_role|>System<|end_role|>
You are a helpful assistant responsible for generating widget schemas that will be used by Instana's widget creation tool to display widgets on their dashboard. Instana is an observability platform that provides real-time insights into the performance and health of applications, services and infrastructure. For the query given to you, you need to return a single word describing the widget type. The return output should be a single word from given options : {"bigNumber", "TIME_SERIES", "pie", "slo2", "topList", "Null"}.If widget type is not provided in the query, you should return "Null".Below is a JSON containing descriptions and the allowed values for widget types, metric names, their corresponding aggregations, and filter options:
{
    "type": {
        "bigNumber": {
            "description": "Displays a single large number in the widget"
        },
        "TIME_SERIES": {
            "description": "Displays data in a time series graph format",
            "type": "TIME_SERIES",
            "formatterSelected": false
        },
        "pie": {
            "description": "Represents data in a pie chart format"
        },
        "slo2": {
            "description": "Gives SLO information"
        },
        "topList": {
            "description": "Lists top N entities for a metric"
        },
        "Null": {
            "description": "No widget type is mentioned"
        }
    }
}
A few examples of queries and their expected outputs are given below
USER: A time series graph for the qotd-author service of the qotd irl application
ASSISTANT: TIME_SERIES
USER: A pie chart for the qotd-author service of the qotd
ASSISTANT: pie
USER: A big number widget for the qotd-author service of the qotd irl application
ASSISTANT: bigNumber
USER: Show the slo configuration for c4ba-macro application
ASSISTANT: slo2
USER: Display the 25th percentile latency trend in the Robot Shop application
ASSISTANT: TIME_SERIES
USER: Show the minimum latency over time for the jdbc service
ASSISTANT: TIME_SERIES
USER: What is the sum of correct calls made to the the GET images endpoint?
ASSISTANT: Null
USER: What is the total number of calls to sli_configurations endpoint in the selfhost1-instana application?
ASSISTANT: bigNumber
USER: Show the mean numeric error rate of erroneous RPC calls to the ratings service
ASSISTANT: bigNumber
USER: Sum of erroneous calls made to catalogue service
ASSISTANT: Null
USER: What is the erroneous call rate per second for PHP Runtime Platform?
ASSISTANT: bigNumber
USER: Show the top 10 endpoints by calls in the Robot Shop application
ASSISTANT: topList
USER: Show the top 5 services with the least latency in the Think Demo Application 
ASSISTANT: topList
USER: Give me the sum of calls 
ASSISTANT: Null<|end_of_text|>
<|start_role|>user<|end_role|>
Query: {}<|end_of_text|>
<|start_role|>assistant<|end_role|>
\end{lstlisting}
\end{longlisting}

\begin{longlisting}{Prompt for Time Series, Big Number and Top List Widget}
\begin{lstlisting}[breaklines=true]
<|start_role|>system<|end_role|> 
You are a helpful assistant responsible for generating widget schemas that will be used by Instana's widget creation tool to display widgets on their dashboard.  Instana is an observability platform provides real-time insights into the performance and health of applications, services and infrastructure. For the query given to you, you need to return a JSON object containing the values for the widget type, metric name, aggregation type of said metric, and filter options.  If any of the values are not provided in the query, you should return "Null" for that value in the JSON object. The return output should be in the following format:
{{
    "type": "widget type (possible values: "bigNumber", "TIME_SERIES", "pie", "topList". If not provided in the query, return "Null")",
    "metric": "metric name (possible values: "calls", "latency", "erroneousCalls", "errors". If not provided in the query, return "Null")",
    "aggregation": "aggregation type (possible values: "SUM", "PER_SECOND", "MEAN", "MIN", "MAX", "P25", "P50", "P75", "P90", "P95", "P98", "P99". If not provided in the query, return "Null")",
    "filter": {{
        "service.name": "service name", "application.name": "application name", "endpoint.name": "endpoint name", "technology.name": "technology name", "call.type": "call type (possible values: "BATCH", "DATABASE", "HTTP", "GRAPHQL", "RPC", "MESSAGING", "OPENTELEMETRY")", "call.erroneous": "whether the call is erroneous (possible values: "true", "false")"
    }}
    "grouping": {{
        "groupbyTag": "Tag used to group results (possible values: "call.error.message", "endpoint.name", "call.http.path", "call.http.status", "http.url", "service.name". If not provided in the query, return "Null")", "direction": "Sort direction for top/bottom highest/lowest selection (possible values: "ASC", "DESC". If not provided in the query, return "Null")", "maxResults": "Maximum number of items to return (possible values: "5", "10", "20", "50". If not provided in the query, return "Null")"
    }}
}}
Your output has to conform to the following under every circumstance without fail:  1. "type", "metric" and "aggregation" are absolutely mandatory and should always be present in the output JSON. Use "Null" if the value is not provided in the query. 
2. Filter options are optional and any one or more of then can be entirely omitted if not provided in the query. You don't have to output "Null" against any particular filter option if it is not present. 
3. Grouping is only application for TIME_SERIES and topList widget types. Any other widgets will NOT have grouping in them. Moreover, grouping is MANDATORY for topList while it is optional for TIME_SERIES. 
4. If a TIME_SERIES widget doesn't have grouping in the query, then don't output the "grouping" field for it. If a topList widget doesn't have grouping in the query, then output "grouping" with "Null" in the corresponding subfields under "grouping". Below is a JSON containing descriptions and the allowed values for widget types, metric names, their corresponding aggregations, and filter options 
<global_knowledge_base> 
Here are some examples:
USER: What is the call rate for the ratings service? Give the exact number
ASSISTANT:{{
    "type": "bigNumber",
    "metric": "calls",
    "aggregation": "PER_SECOND",
    "filter": {{
        "service.name": "ratings"
    }}
}}
USER: Display the per-second rate of erroneous batch calls in the Robot Shop application as a single number
ASSISTANT:{{
    "type": "bigNumber",
    "metric": "erroneousCalls",
    "aggregation": "PER_SECOND",
    "filter": {{
        "application.name": "Robot Shop",
        "call.type": "BATCH"
    }}
}}
USER: What is the sum of erroneous calls made to the catalogue service over time? 
ASSISTANT:{{
    "type": "TIME_SERIES",
    "metric": "erroneousCalls",
    "aggregation": "SUM",
    "filter": {{
        "service.name": "catalogue"
    }}
}}
USER: Display the total number of erroneous HTTP calls made to the user service over time
ASSISTANT:{{
    "type": "TIME_SERIES",
    "metric": "erroneousCalls",
    "aggregation": "SUM",
    "filter": {{
        "service.name": "user",
        "call.type": "HTTP"
    }}
}}
USER: Display the per-second erroneous call rate to the POST /login endpoint at any given moment of time
ASSISTANT:{{
    "type": "TIME_SERIES",
    "metric": "erroneousCalls",
    "aggregation": "PER_SECOND",
    "filter": {{
        "endpoint.name": "POST /login"
    }}
}}
USER: Show the 90th percentile latency trend of correct calls to the user service
ASSISTANT:{{
    "type": "TIME_SERIES",
    "metric": "latency",
    "aggregation": "P90",
    "filter": {{
        "service.name": "user",
        "call.erroneous": "false"
    }}
}}
USER: Erroneous calls made per second to the nginx-web service of the robot shop application
ASSISTANT:{{
    "type": "null",
    "metric": "erroneousCalls",
    "aggregation": "PER_SECOND",
    "filter": {{
        "application.name": "Robot Shop",
        "service.name": "nginx-web"
    }}
}}
USER: Big number widget for the qotd-engraving service
ASSISTANT:{{
    "type": "bigNumber",
    "metric": "Null",
    "aggregation": "Null",
    "filter": {{
        "service.name": "qotd-engraving"
    }}
}}
USER: What is mean error rate over time?
ASSISTANT:{{
    "type": "TIME_SERIES",
    "metric": "errors",
    "aggregation": "MEAN",
    "filter": {{}}
}}
USER: Display the calls per second over time for the 'Product X' application
ASSISTANT:{{
    "type": "TIME_SERIES",
    "metric": "calls",
    "aggregation": "PER_SECOND",
    "filter": {{
        "application.name": "Product X"
    }}
}}
USER: A time series graph for the qotd-author service of the qotd irl application
ASSISTANT:{{
    "type": "TIME_SERIES",
    "metric": "Null",
    "aggregation": "Null",
    "filter": {{
        "application.name": "qotd irl",
        "service.name": "qotd-author"
    }}
}}
USER: Show the top 10 endpoints by highest sum of calls in the Robot Shop application
ASSISTANT:{{
  "type": "topList",
  "metric": "calls",
  "aggregation": "SUM",
  "filter": {{
    "application.name": "Robot Shop"
  }},
  "grouping": {{
    "groupbyTag": "endpoint.name",
    "direction": "DESC",
    "maxResults": "10"
  }}
}}
USER: Give me the top 5 services of the Think Demo 2025 Application with the lowest mean latency
ASSISTANT: {{
  "type": "topList",
  "metric": "latency",
  "aggregation": "MEAN",
  "filter": {{
    "application.name": "Think Demo 2025"
  }},
  "grouping": {{
    "groupbyTag": "service.name",
    "direction": "ASC",
    "maxResults": "5"
  }}
}}
USER: Give me the top endpoints with highest latency 
ASSISTANT: {{
  "type": "topList",
  "metric": "latency",
  "aggregation": "Null",
  "filter": {{}},
  "grouping": {{
    "groupbyTag": "endpoint.name",
    "direction": "DESC",
    "maxResults": "Null"
  }}
}}
USER: Bottom services in the Think demo 2025 application
ASSISTANT: {{
  "type": "topList",
  "metric": "Null",
  "aggregation": "Null",
  "filter": {{
    "application.name": "Think demo 2025"
  }},
  "grouping": {{
    "groupbyTag": "service.name",
    "direction": "ASC",
    "maxResults": "Null"
  }}
}}<|end_of_text|>
<|start_role|>user<|end_role|>
Query: {}<|end_of_text|>
<|start_role|>assistant<|end_role|>
\end{lstlisting}
\end{longlisting}

\begin{longlisting}{Prompt for SLO Widget}
\begin{lstlisting}
<|start_role|>system<|end_role|> 
You are a helpful assistant responsible for generating widget schemas that will be used by Instana's widget creation tool to display widgets on their dashboard. Instana is an observability platform provides real-time insights into the performance and health of applications, services and infrastructure. For the query given to you, you need to return a JSON object that indicates the name of the SLO configuration requested by the user. If SLO configuration is not provided in the query, you should return "None". Some examples are given below:
USER: Generate an SLO widget for the SLO configuration Great Expectations 
ASSISTANT: {{"name": "Great Expectations"}}
USER: Generate an SLO widget for the SLO configuration [demo]RobotShop - ep 123
ASSISTANT: {{"name": "[demo]RobotShop - ep 123"}}<|end_of_text|>
<|start_role|>user<|end_role|>
Query: {}<|end_of_text|>
<|start_role|>assistant<|end_role|>
\end{lstlisting}
\end{longlisting}

\subsection{B. Error Analysis}

\begin{table*}[htbp]
\centering

\begin{tabular}{|l|c|c|c|}
\hline
\textbf{Model} & \textbf{TagFilter (Failures \textbar{} Error\%)} & \textbf{Grouping (Failures \textbar{} Error\%)} & \textbf{Samples (TagF \textbar{} Group)} \\
\hline
llama-3-3-70b & 65 \textbar{} 25.79\% & 17 \textbar{} \textbf{35.42\%} & 252 \textbar{} 48 \\
gpt-oss-120b & 65 \textbar{} 25.79\% & 17 \textbar{} \textbf{35.42\%} & 252 \textbar{} 48 \\
gpt-oss-20b & 65 \textbar{} 25.79\% & 17 \textbar{} \textbf{35.42\%} & 252 \textbar{} 48 \\
granite-3-2b & 66 \textbar{} 26.19\% & 17 \textbar{} \textbf{35.42\%} & 252 \textbar{} 48 \\
llama-4-maverick-17b-128e & 54 \textbar{} \textbf{21.43\%} & 21 \textbar{} 43.75\% & 252 \textbar{} 48 \\
Mistral-Small-3.1-24B & 55 \textbar{} \textbf{21.83}\% & 21 \textbar{} 43.75\% & 252 \textbar{} 48 \\
\hline
\end{tabular}
\caption{Performance comparison of different models on TagFilter and Grouping tasks.}
\label{tab:model_failures}
\end{table*}

\begin{table*}[t]
\centering
\renewcommand{\arraystretch}{1.2}
\begin{tabularx}{\textwidth}{p{0.25\textwidth} p{0.18\textwidth} X}
\hline
\textbf{Prompt} & \textbf{Uniquely Correct Model} & \textbf{Errors by Other Models} \\
\hline
Display the per-second erroneous call rate to the \texttt{POST /login} endpoint at any given moment of time 
& llama-3-3-70b-instruct 
& All others predicted \texttt{bigNumber} instead of correct \texttt{time series}. Tag filter was mostly correct, but incorrect widget type choice caused overall failure. \\

What is the sum of erroneous HTTP calls made to the user service 
& gpt-oss-20b 
& All other models predicted \texttt{bigNumber} correctly but missed subtle handling of null widgetType or over-specified query, leading to overall misclassification. \\

Display the 99th percentile OpenTelemetry latency value for erroneous calls to \texttt{otel-shop-cart} service 
& llama-4-maverick-17b-128e-instruct-fp8 
& All others either missed the \texttt{call.type=OPENTELEMETRY} filter or predicted the wrong widget type (\texttt{time series} vs. required \texttt{bigNumber}). Incorrect tag filter specification led to failure. \\
\hline
\end{tabularx}
\caption{Representative failure cases where only one model produced the correct output while all others failed}
\label{tab:unique-failures}
\end{table*}

A closer examination of the results reveals that the main difficulties lie in interpreting nuanced and underspecified aspects of the user's query, specifically in Tag Filter Expression and Grouping. Table~\ref{tab:model_failures} gives a statistical view of these errors. Table~\ref{tab:unique-failures} shows representative failure cases where only one model produced the correct output while all others failed. Each row highlights the query, the uniquely correct model, and the systematic errors made by the other models

\textbf{Tag Filter Expression Errors.} While achieving a respectable accuracy of up to $80.07\%$, the Tag Filter Expression extraction still presents a challenge. Errors in this category typically arise from:
\begin{itemize}
    \item \textit{Ambiguous Naming:} This is the most common type of error, where the model incorrectly identifies an entity or maps it to the wrong tag. This often occurs when there are subtle differences in naming, such as the model predicting \texttt{`payment'} instead of the correct \texttt{`otel-shop-payment'}. These errors highlight the challenges in fuzzy entity matching and the need for a more robust entity-linking mechanism.
    \item \textit{Incomplete Extraction:} In these cases, the model fails to extract all the necessary tags from the query. For example, when asked for \texttt{`HTTP calls made to the payment service'}, the model might correctly identify the service but omit the \texttt{call.type: `HTTP' tag}. This results in a widget that is overly broad and does not meet the user's specific request.
    \item \textit{Complete Failure of Extraction:} Less frequently, the model fails to extract any tag filter expression, even when the query contains filtering criteria. This often happens with more complex or multi-part queries, where the model seems to lose context.
\end{itemize}

\textbf{Grouping Errors.} The extraction of Grouping parameters presents its own set of challenges, primarily related to the interpretation of user intent. The errors in this category are most often due to:
\begin{itemize}
    \item \textit{Implicit Parameter Extraction:} The most frequent error in this category is the failure to extract implicit parameters, especially \texttt{maxResults}. Users often phrase their requests as ``top 5" or ``show me the 10 services," but the model fails to translate this into the \texttt{maxResults} parameter in the grouping expression. This suggests a limitation in the model's ability to reason about the query beyond a direct, literal interpretation.
    \item \textit{Incorrect Grouping Tag:} In some cases, the model selects the wrong tag for grouping. For example, a user might ask for ``the latency of services," and the model might incorrectly group by \texttt{application.name} instead of \texttt{service.name}. This type of error points to a misunderstanding of the underlying data schema and the relationships between different entities.
    \item \textit{Underspecified Queries:} Many queries are inherently ambiguous when it comes to grouping. A user might ask for \textit{``the services with the highest latency"} which implies a descending order, but the model may not always infer this correctly. This is a key area where the proposed multi-turn clarification loop can play a crucial role. By asking the user to clarify their intent, NOVAID can resolve this ambiguity and generate the correct widget.
\end{itemize}

\subsection{C. Deployment Challenges}
Deploying NOVAID in real enterprise ITOps environments introduced several practical challenges that required engineering solutions beyond the core research contributions. Below, we summarize key deployment hurdles and how they were addressed.

\textbf{Prompt Size and Token Limitations.} A natural approach was to embed all service names, application names, and endpoint names directly into the LLM prompt. However, this caused the prompt size to grow prohibitively large and occasionally exceed model token limits. To mitigate this, we adopted a two-stage strategy: the LLM first identified the high-level intent and the target filter category using only its foundational knowledge, after which a cosine similarity–based fuzzy matching algorithm determined the exact service, application, or endpoint name from the tenant-specific entity set. This reduced prompt length while retaining semantic accuracy in entity identification.

\textbf{Tenant-Specific and Dynamic Entity Names.}
In production deployments, entity names (e.g., services, endpoints) are highly dynamic and differ across tenant environments. We resolved this by fetching entity names in real time using each tenant’s authentication key, passed within the body of the backend API call to Instana APIs. The system architecture was designed modularly so that retrieved entities were cached locally and could be reloaded automatically. This ensured NOVAID could seamlessly adapt to new tenants without code changes.

\textbf{Entity Drift and Latency Overheads.}
Since entities evolve during deployment (e.g., new services, renamed endpoints), periodic refreshing was necessary. Frequent refetching, however, created latency overheads. To find an optimal balance, we conducted a survey with IT professionals to understand realistic expectations of refresh frequency. Based on this, we scheduled refetching at user-validated intervals, thereby reducing unnecessary overhead without compromising usability. Furthermore, we introduced parallel multithreaded asynchronous requests for entity retrieval. This optimization reduced fetching latency from $\sim$8 seconds to $\sim$2 seconds in our test environment, significantly improving responsiveness.

\textbf{User Uncertainty in Service Selection.}
In practice, users were often unsure of the exact service name to query and would select multiple candidates before identifying the correct widget from the preview. This behavior created excessive backend calls. To address this, we kept all the modules decoupled from each other and loosely connected. The fuzzy matching module generated a list of plausible candidate outputs once, and this candidate set was passed to the widget creation tool, which rendered real-time previews. Users could then iteratively refine their choice without triggering additional LLM calls. This design substantially reduced both backend workload and widget rendering latency.

These deployment-focused improvements ensured NOVAID’s robustness in dynamic, multi-tenant environments while maintaining low latency and high user satisfaction. The practical lessons learned from handling prompt size constraints to optimizing entity retrieval pipeline demonstrate the system’s readiness for enterprise deployment and highlight considerations for similar natural language–to-ITOps systems.

\begin{figure*}[htbp!]
    \centering
    \begin{subfigure}{0.48\textwidth}
        \includegraphics[width=\linewidth]{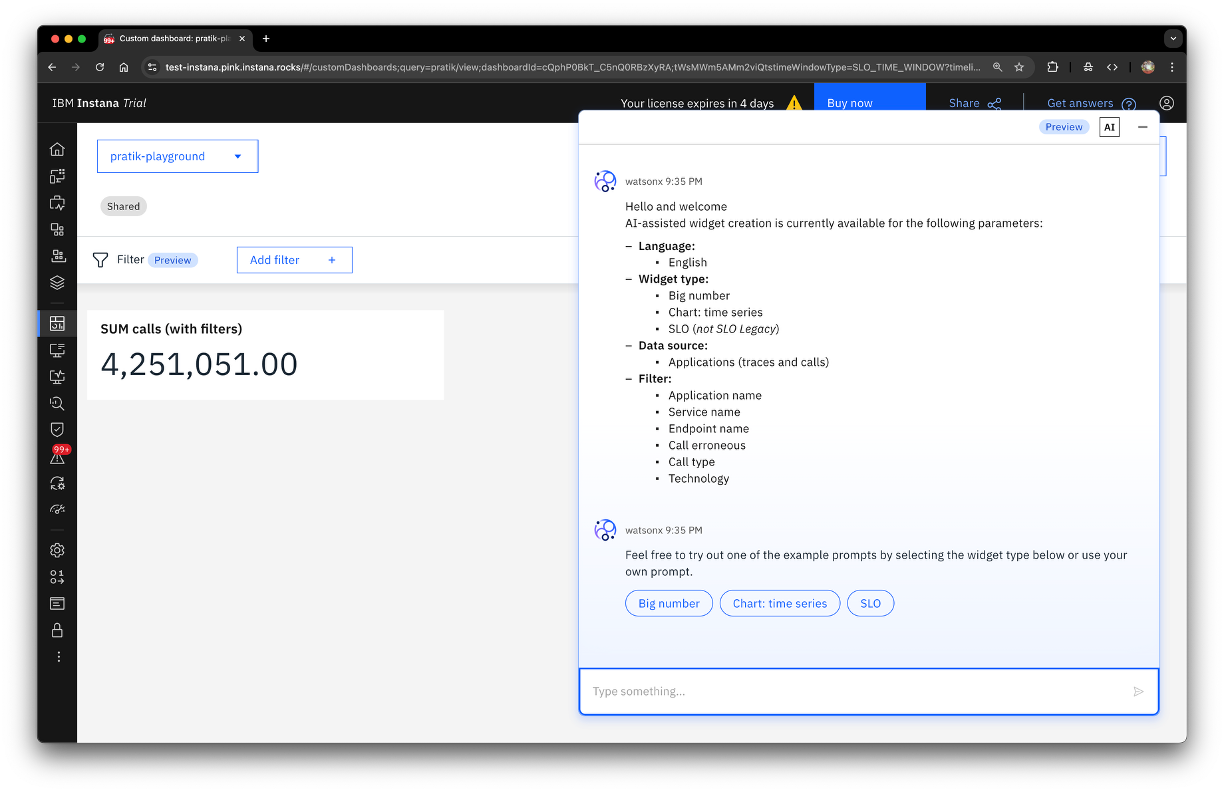}
        \caption{Landing Page}
    \end{subfigure}
    \begin{subfigure}{0.48\textwidth}
        \includegraphics[width=\linewidth]{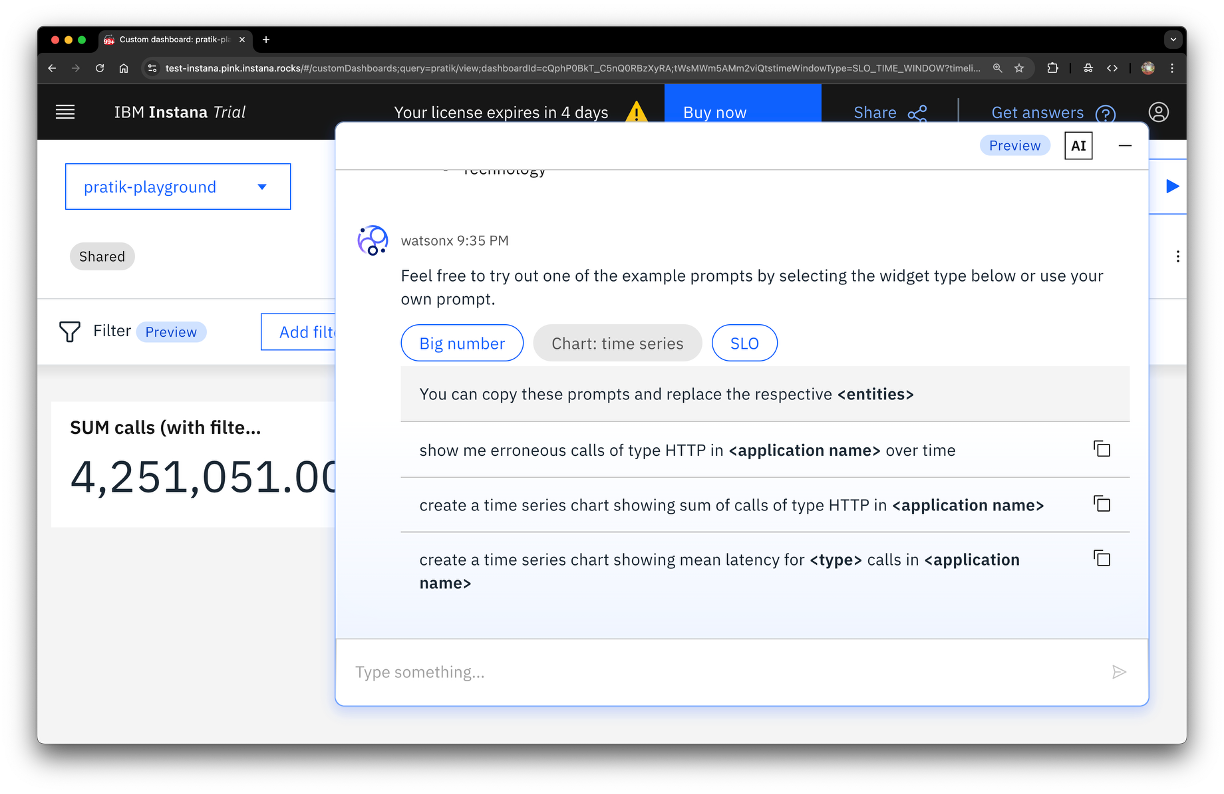}
        \caption{Example Prompts}
    \end{subfigure}
    \begin{subfigure}{0.48\textwidth}
        \includegraphics[width=\linewidth]{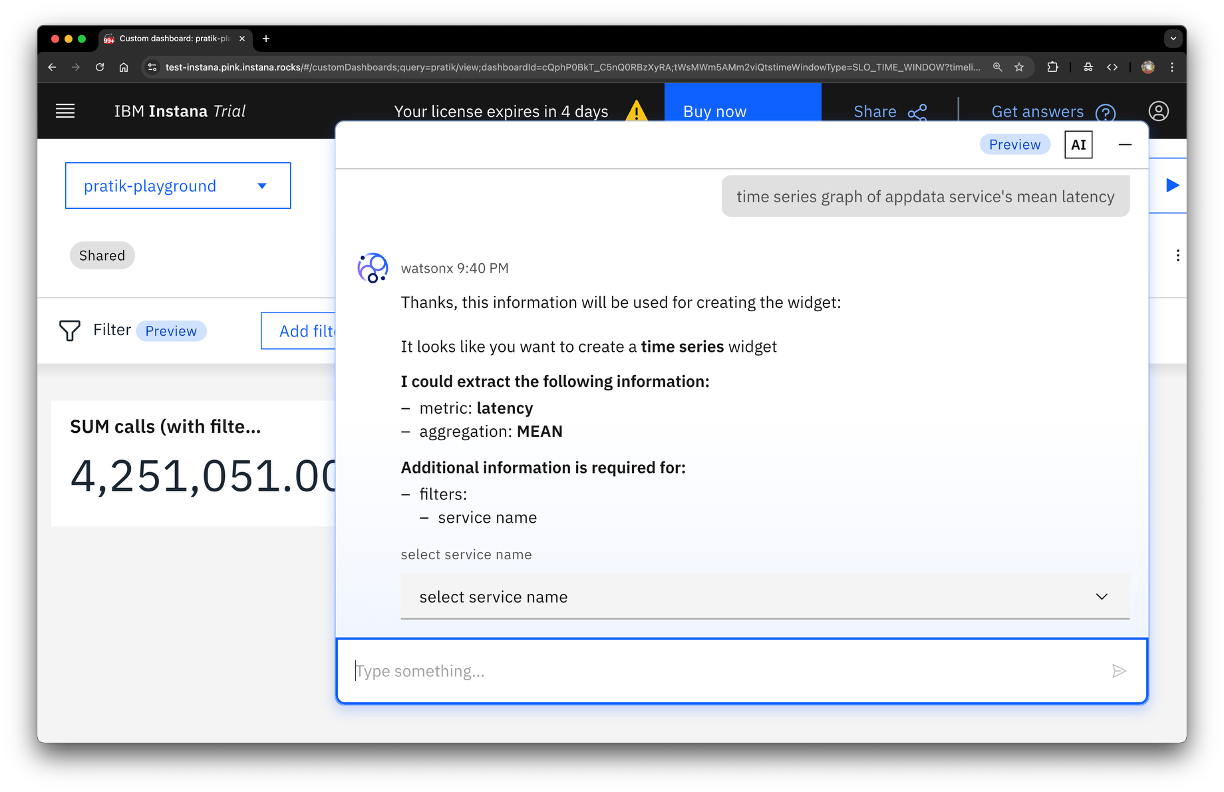}
        \caption{Query Input}
    \end{subfigure}
    \begin{subfigure}{0.48\textwidth}
        \includegraphics[width=\linewidth]{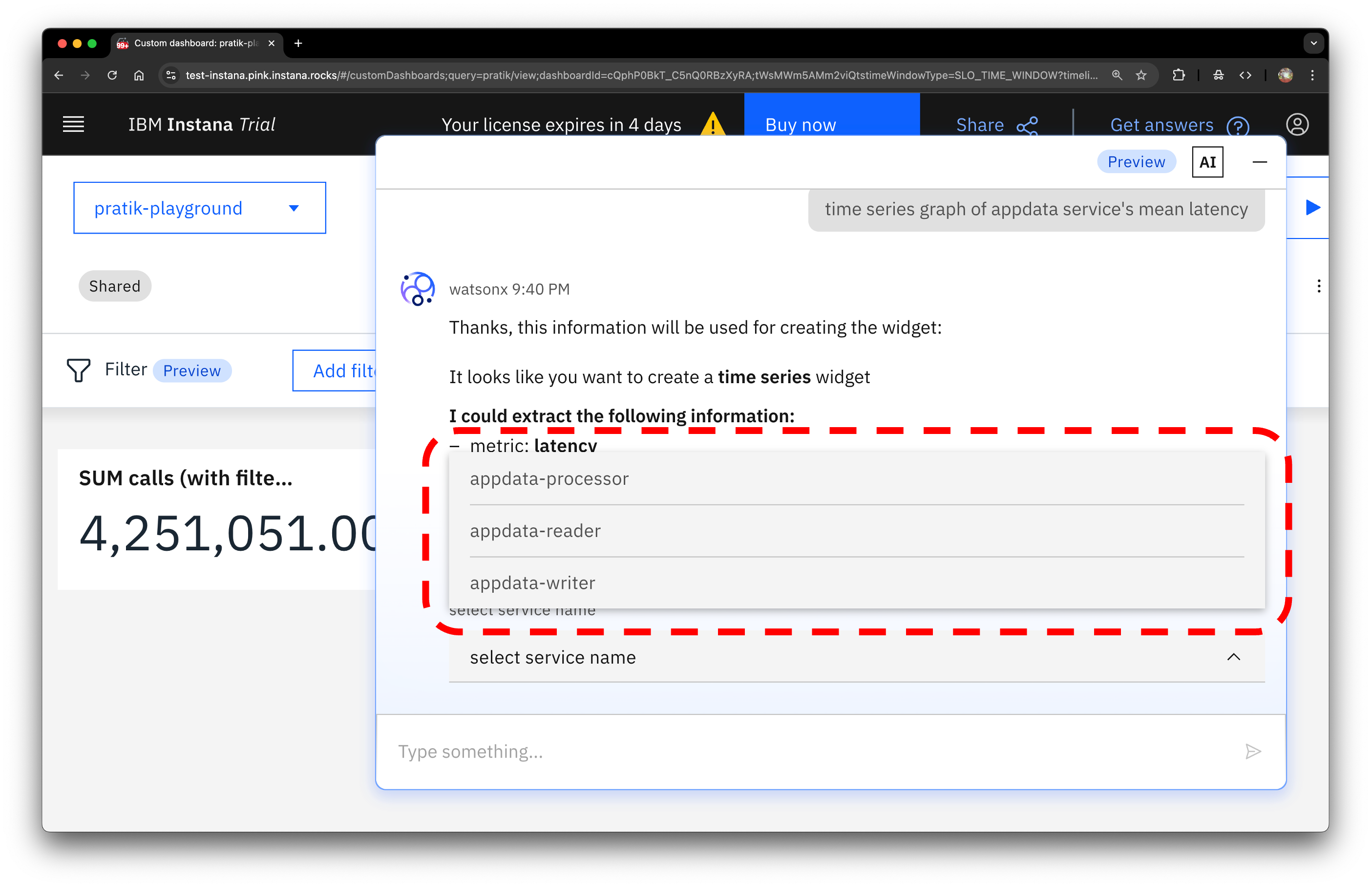}
        \caption{Dropdown Selection}
    \end{subfigure}
    \begin{subfigure}{0.48\textwidth}
        \includegraphics[width=\linewidth]{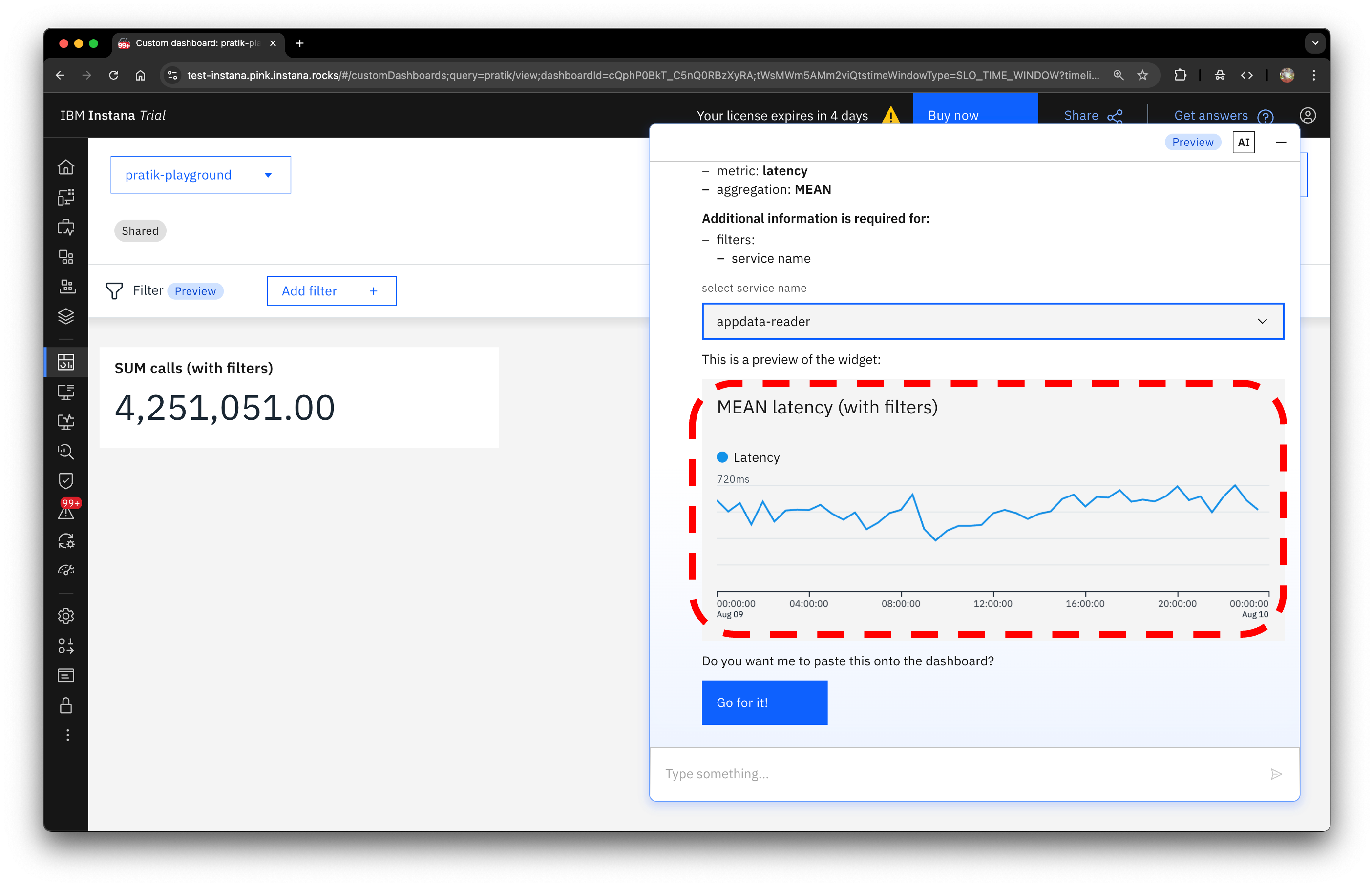}
        \caption{Preview Window}
    \end{subfigure}
    \begin{subfigure}{0.48\textwidth}
        \includegraphics[width=\linewidth]{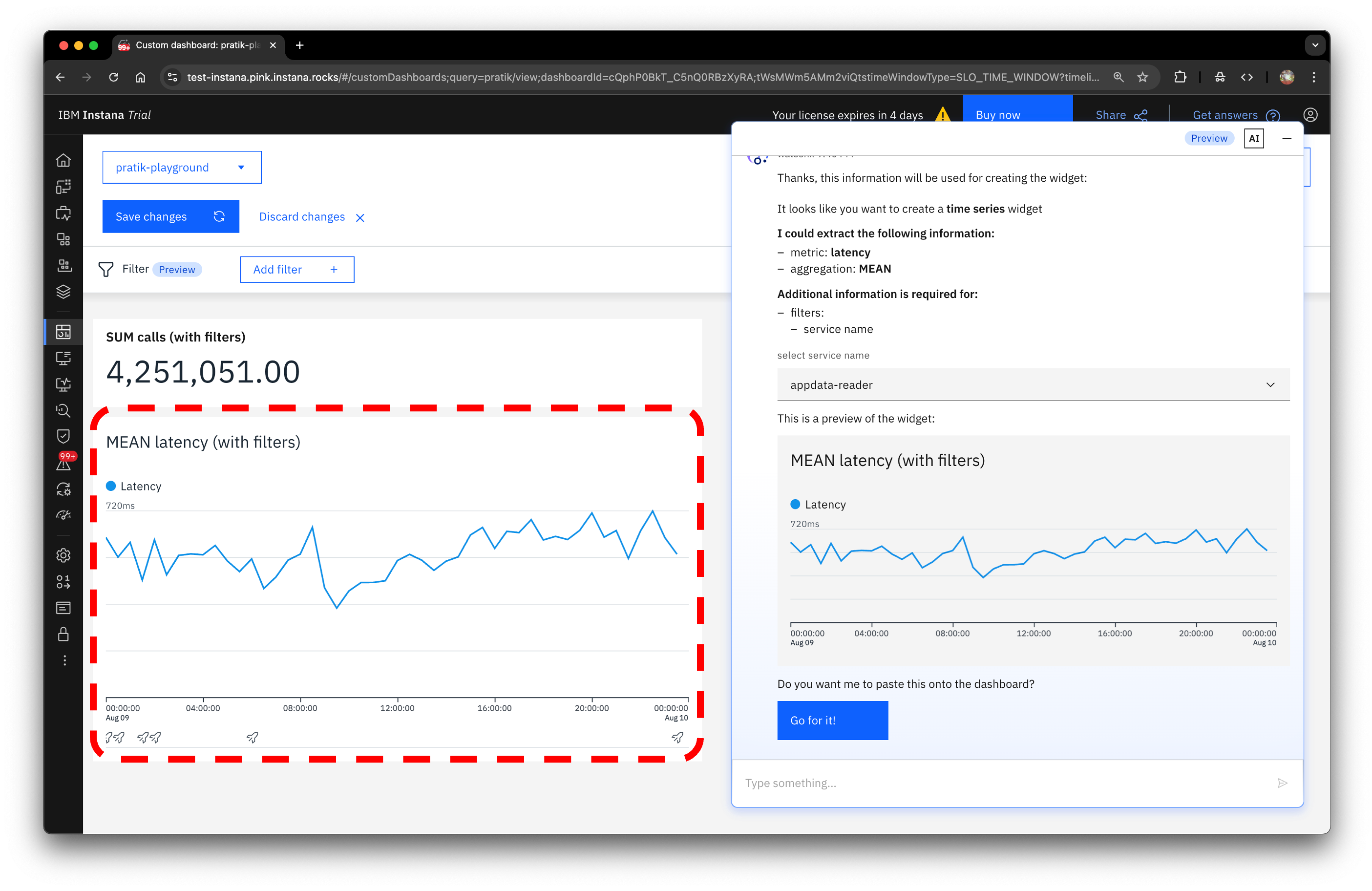}
        \caption{Final Dashboard}
    \end{subfigure}
    
    \caption{User workflow illustration: (a) landing page, (b) example prompts, (c) query input, (d) dropdown selection, (e) widget preview, and (f) dashboard output.}
    \label{fig:user_workflow}
\end{figure*}

\begin{figure*}[htbp!]
    \centering
    \begin{subfigure}{0.33\textwidth}
        \includegraphics[width=\linewidth]{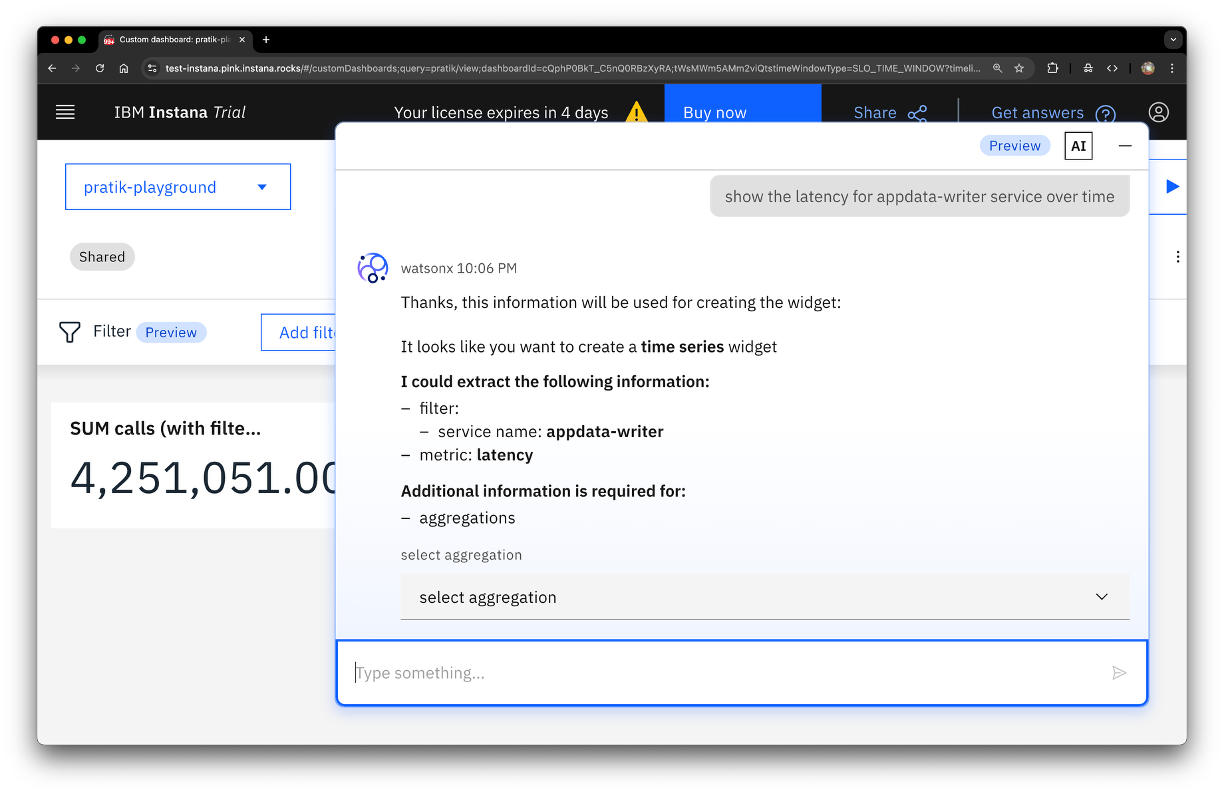}
        \caption{Query Input}
    \end{subfigure}
    \begin{subfigure}{0.33\textwidth}
        \includegraphics[width=\linewidth]{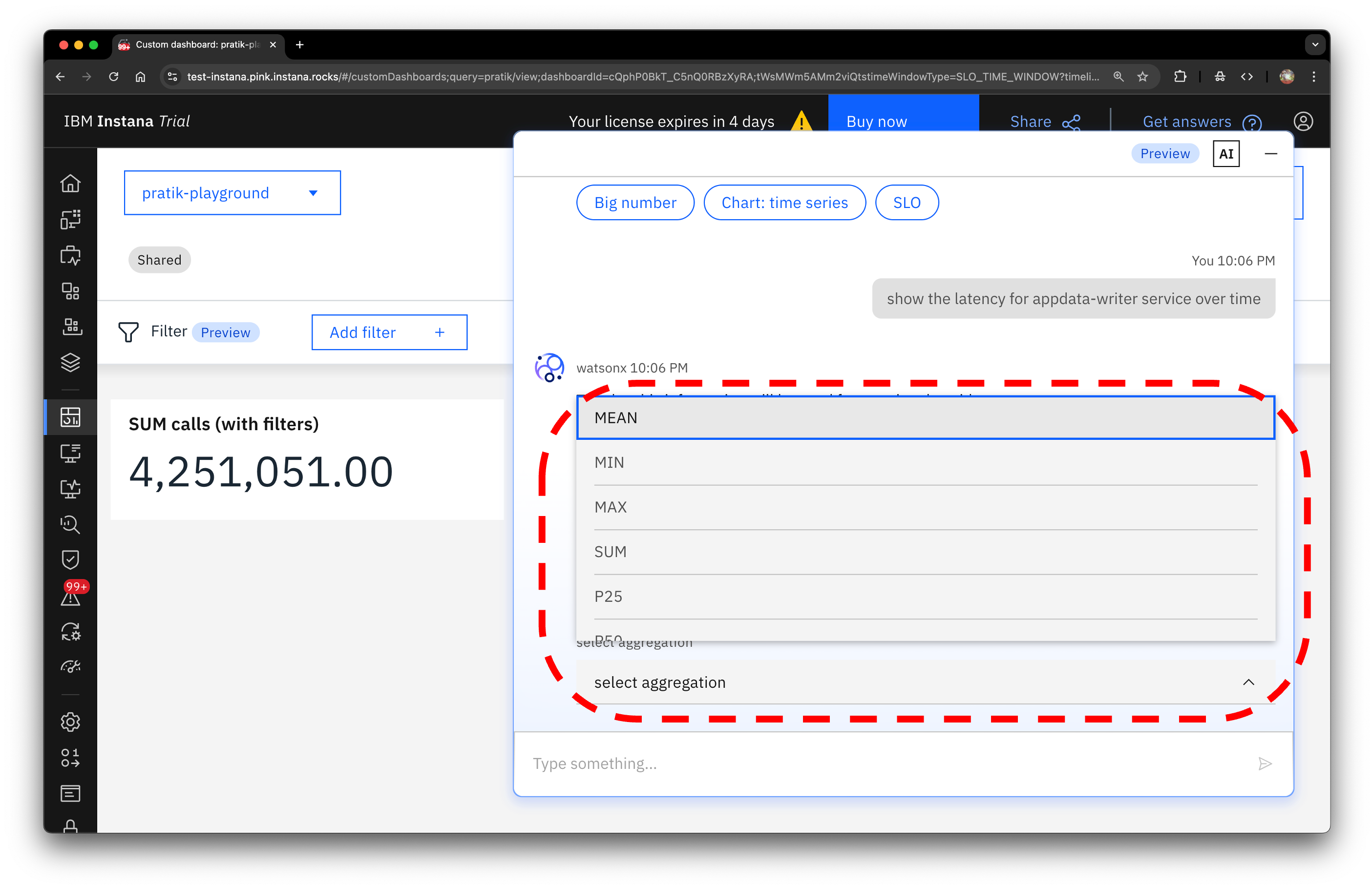}
        \caption{Dropdown Selection}
    \end{subfigure}
    \begin{subfigure}{0.33\textwidth}
        \includegraphics[width=\linewidth]{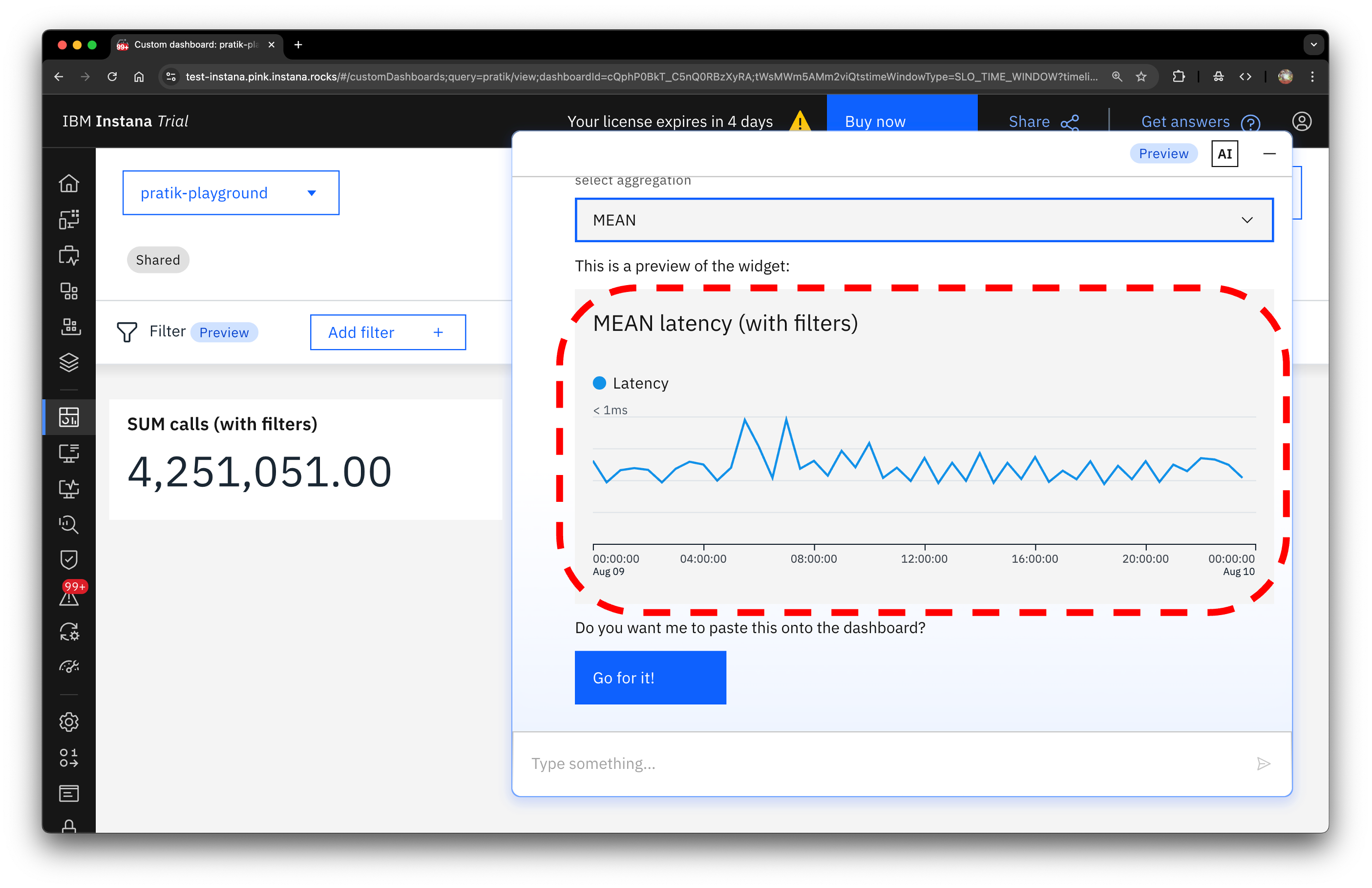}
        \caption{Preview Window}
    \end{subfigure}
    
    \caption{User journeys (a–c) for query ``show the latency for appdata-writer service over time'' where NOVAID infers the widget type as a time series chart even though not explicitly mentioned}
    \label{fig:scenario-2}
\end{figure*}

\begin{figure}[htbp!]
    \includegraphics[width=\linewidth]{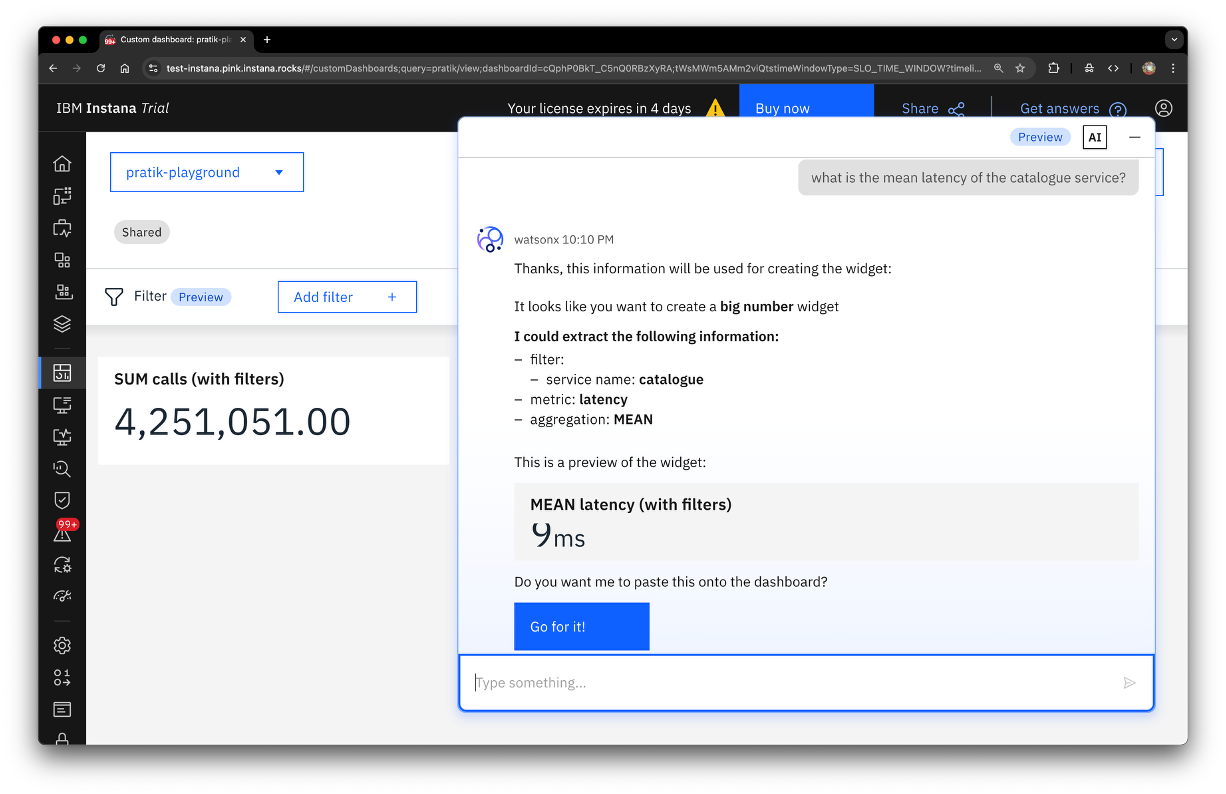}
    \caption{Big Number Widget}
    \label{fig:scenario-3}
\end{figure}

\subsection{D. User Journey Examples}

Figure~\ref{fig:user_workflow} illustrates the user journey with NOVAID's interface.  
The landing page (Figure~\ref{fig:user_workflow}a) comprises a conversational interface on the right and a dashboard canvas on the left. The chat interface provides example prompts for various widget types (Figure~\ref{fig:user_workflow}b), allowing users to experiment with them. Entering a natural language query in the chat box and pressing the ``enter” key or the blue arrow button triggers the system to parse the query.  

For the entered query, \textit{``time series graph of appdata service's mean latency"}, NOVAID infers the widget type, metric name, and aggregation method, and identifies that the service name filter information is missing (Figure~\ref{fig:user_workflow}c). A dropdown menu with relevant service names is then presented (Figure~\ref{fig:user_workflow}d). Once the user chooses a service name, the panel shows a preview of the widget (Figure~\ref{fig:user_workflow}e). Pressing \texttt{Go for it!} moves the widget to the main dashboard canvas (Figure~\ref{fig:user_workflow}f).

The user journey for the query \textit{``show the latency for appdata-writer service over time"} is shown in Figure~\ref{fig:scenario-2}a–c. Here, NOVAID correctly infers the widget type to be a \texttt{"Chart: time series"}, even when it is not explicitly mentioned. Another example is shown in Figure~\ref{fig:scenario-3}, where NOVAID extracts key elements from the query and infers the widget type as a \texttt{"Big number"}.

\subsection{E. User Study Qualitative Feedback}
Below are a few selected subjective comments from users:
\begin{itemize}
\item I think that I would like to use this system frequently. 
\textit{It was easier than going through the documentation on what all settings need to be applied.}

\item I found the system unnecessarily complex. 
\textit{It was not complex. I was rather surprised how good the suggestions in the dropdowns were and how well the AI ironed out any uncertainties by asking questions.}

\item I thought the system was easy to use. 
\textit{I think it could be improved by giving some suggestions for commonly used prompts, so you don't have to start from a blank chat.}

\item I think that I would need the support of a technical person to be able to use this system. 
 \textit{Not really, pretty simple to use, given familiarity with the underlying functionality.}

 \item I thought there was too much inconsistency in this system. \textit{I didn't experience any inconsistency.} 
 \textit{I think since this is a work in progress, there is some issue with finding the correct application/metrics/filter.}

\item I would imagine that most people would learn to use this system very quickly. 
 \textit{Easy with examples of prompts to use.}

 \item I found the system very cumbersome to use. 
 \textit{Easy to use.} \textit{The response was very fast.}

\item I felt very confident using the system. 
\textit{You either get exactly what was requested, or easy to edit to get the results.}

 \item I needed to learn a lot of things before I could get going with this system. 
 \textit{Equipped with underlying knowledge of the chart functionality, not much to learn.} 

\item Using this tool will increase my productivity. \textit{Over the long run I think creating charts could go quicker, might help to be able to ask questions like what metrics/aggregations are available or even what charts.}

\end{itemize}

To summarize, the majority of the participants emphasised the ease of use and effectiveness of NOVAID. Participants mentioned that getting suggestions from the model was helpful to ask more precise queries, thereby making the system more reliable. Regarding areas for improvement, some participants mentioned that having more example utterances would help them better understand the system’s capabilities. A few participants reported NOVAID's challenges in extracting the correct metric names or filters. Others expressed interest in being able to ask specific starting point questions, such as which metrics, aggregations, or widget types are available.

\end{document}